\documentclass [] {aa} 
\usepackage{epsfig, graphicx}

\begin{document}

%\thesaurus{06(08.01.1; 08.03.02; 08.16.2)}

\title{Nitrogen abundances in Planet-harbouring stars}

\author{A.~Ecuvillon\inst{1}, G.~Israelian\inst{1}, N. C.~ Santos\inst{2,3}, 
M.~Mayor\inst{3}, R. J.~Garc\'{\i}a L\'opez\inst{1,4} \and S.~Randich\inst{5}}

\offprints{ \email{aecuvill@ll.iac.es}}

\institute{Instituto de Astrof\'{\i}sica de Canarias, E-38200 La Laguna, Tenerife, Spain \and Centro de 
Astronomia e Astrofisica de Universidade de Lisboa, Observatorio Astronomico de Lisboa, Tapada de Ajuda, 1349-018 Lisboa, 
Portugal \and Observatoire de Gen\`eve, 51 ch.  des  Maillettes, CH--1290 Sauverny, Switzerland \and Departamento de 
Astrof\'{\i}sica, Universidad de La Laguna, Av. Astrof\'{\i}sico Francisco S\'anchez s/n, E-38206 La Laguna, Tenerife, 
Spain \and INAF/Osservatorio Astrofisico di Arcetri, Largo Fermi 5, 50125 Firenze, Italy }

\date{Received 21 November 2003/ Accepted 19 January 2004} 

\titlerunning{Nitrogen abundances in Planet-harbouring stars} 
\authorrunning{A. Ecuvillon et al.}
%--------------------------------------------------------------------------

\abstract{
We present a detailed spectroscopic analysis of nitrogen abundances in 91
solar-type stars, 66 with and 25 without known planetary mass companions.
All comparison sample stars and 28 planet hosts were analysed by spectral synthesis
of the near-UV NH band at 3360 \AA\ observed at high resolution with the VLT/UVES,
while the near-IR N\,{\sc i} 7468 \AA\ was measured in 31 objects. These two 
abundance indicators are in good agreement. We found that nitrogen abundance scales
with that of iron in the metallicity range $-0.6 <$[Fe/H]$<+0.4$ with the slope
$1.08 \pm 0.05$. Our results show that the bulk of nitrogen production at high
metallicities was coupled with iron. We found that the nitrogen abundance
distribution in stars with exoplanets is the high [Fe/H] extension of the curve
traced by the comparison sample of stars with no known planets. A comparison of our 
nitrogen abundances with those available in the literature shows a good agreement. 
	  }
\maketitle

\section{Introduction}
\label{Intro}

\begin{table*}[t]
\caption[]{Observing log for the new set of stars with planets and brown dwarf companions. Near-UV and 
optical spectra were 
obtained with UVES, and with SARG and FEROS, respectively. The S/N ratio is provided at 3357 \AA\ for near-UV 
data and at 7400 \AA\ 
in optical spectra.}
\begin{center}
\begin{scriptsize}
\begin{tabular}{lcccr}
\hline
\noalign{\smallskip}
Star & $V$ & Obser. Run & S/N & Date \\
\hline
\hline
\noalign{\smallskip}
\object{HD\,6434}   & 7.7 & UVES  & 150 & Oct.\ 2001 \\
\object{HD\,22049}  & 3.7 & UVES  & 150  & Oct.\ 2001 \\
\object{HD\,30177}  & 8.4 & FEROS & 200 & Mar.\ 2003 \\
\object{HD\,40979}  & 6.8 & SARG  & 220 & Oct.\ 2003 \\
\object{HD\,46375}  & 7.8 & UVES  & 150 & Nov.\ 2001 \\
\object{HD\,65216}  & 9.6 & FEROS & 250 & Mar.\ 2003 \\
\object{HD\,68988}  & 8.2 & SARG  & 160 & Oct.\ 2003 \\
\object{HD\,72659}  & 7.4 & FEROS & 200 & Mar.\ 2003 \\
\object{HD\,73256}  & 8.1 & FEROS & 200 & Mar.\ 2003 \\
\object{HD\,73526}  & 9.0 & FEROS & 250 & Mar.\ 2003 \\
\object{HD\,76700}  & 8.1 & FEROS & 200 & Mar.\ 2003 \\
\object{HD\,83443}  & 8.2 & UVES  & 120 & Nov.\ 2001 \\
\object{HD\,10647}  & 5.5 & UVES  & 160 & Oct.\ 2001 \\
\object{HD\,142415} & 7.3 & FEROS & 250 & Mar.\ 2003 \\
\object{HD\,169830} & 5.9 & UVES  & 160 & Oct.\ 2001 \\
\object{HD\,178911B}& 6.7 & SARG  & 280 & Oct.\ 2003 \\
\object{HD\,179949} & 6.3 & UVES  & 160 & Oct.\ 2001 \\
\object{HD\,202206} & 8.1 & UVES  & 150 & Nov.\ 2001 \\
\object{HD\,209458} & 7.7 & UVES  & 150 & Nov.\ 2001 \\
\object{HD\,216770} & 8.1 & SARG  & 220 & Oct.\ 2003 \\
\object{HD\,219542B}& 8.2 & SARG  & 230 & Oct.\ 2003 \\
\object{HD\,222582} & 7.7 & UVES  & 180 & Nov.\ 2001 \\
\noalign{\smallskip}
\hline
\end{tabular}
\end{scriptsize}
\end{center}
\label{tab1}
\end{table*}

\begin{table*}[t]
\caption[]{Observing log for the new set of comparison stars (stars without known giant planets). Near-UV spectra are
obtained with UVES. The S/N ratio is provided at 3357 \AA.}
\begin{center}
\begin{scriptsize}
\begin{tabular}{lcccccr}
\hline
\noalign{\smallskip}
Star & $V$ & Obser. Run & S/N & Date \\
\hline 
\hline
\noalign{\smallskip}
\object{HD\,4391}    & 5.8 & UVES & 240 & Oct.\ 2001 \\
\object{HD\,7570}    & 4.9 & UVES & 150 & Oct.\ 2001 \\
\object{HD\,10700}   & 3.5 & UVES & 200 & Oct.\ 2001 \\
\object{HD\,14412}   & 6.3 & UVES & 130 & Oct.\ 2001 \\
\object{HD\,20010}   & 3.9 & UVES & 240 & Oct.\ 2001 \\
\object{HD\,20766}   & 5.5 & UVES & 140 & Oct.\ 2001 \\
\object{HD\,20794}   & 4.3 & UVES & 130 & Oct.\ 2001 \\
\object{HD\,20807}   & 5.2 & UVES & 180 & Oct.\ 2001 \\
\object{HD\,23484}   & 7.0 & UVES & 160 & Oct.\ 2001 \\
\object{HD\,30495}   & 5.5 & UVES & 120 & Oct.\ 2001 \\
\object{HD\,36435}   & 7.0 & UVES & 110 & Oct.\ 2001 \\
\object{HD\,38858}   & 6.0 & UVES &  80 & Oct.\ 2001 \\
\object{HD\,43162}   & 6.4 & UVES & 180 & Nov.\ 2001 \\
\object{HD\,43834}   & 5.1 & UVES & 180 & Nov.\ 2001 \\
\object{HD\,69830}   & 5.9 & UVES & 130 & Oct.\ 2001 \\
\object{HD\,72673}   & 6.4 & UVES & 110 & Nov.\ 2001 \\
\object{HD\,76151}   & 6.0 & UVES &  90 & Oct.\ 2001 \\
\object{HD\,84117}   & 4.9 & UVES & 200 & Nov.\ 2001 \\
\object{HD\,189567}  & 6.1 & UVES & 140 & Oct.\ 2001 \\
\object{HD\,192310}  & 5.7 & UVES &  90 & Oct.\ 2001 \\
\object{HD\,211415}  & 5.3 & UVES & 120 & Oct.\ 2001 \\
\noalign{\smallskip}
\hline
\end{tabular}
\end{scriptsize}
\end{center}
\label{tab2}
\end{table*}

\begin{figure*}
\includegraphics[height=6cm]{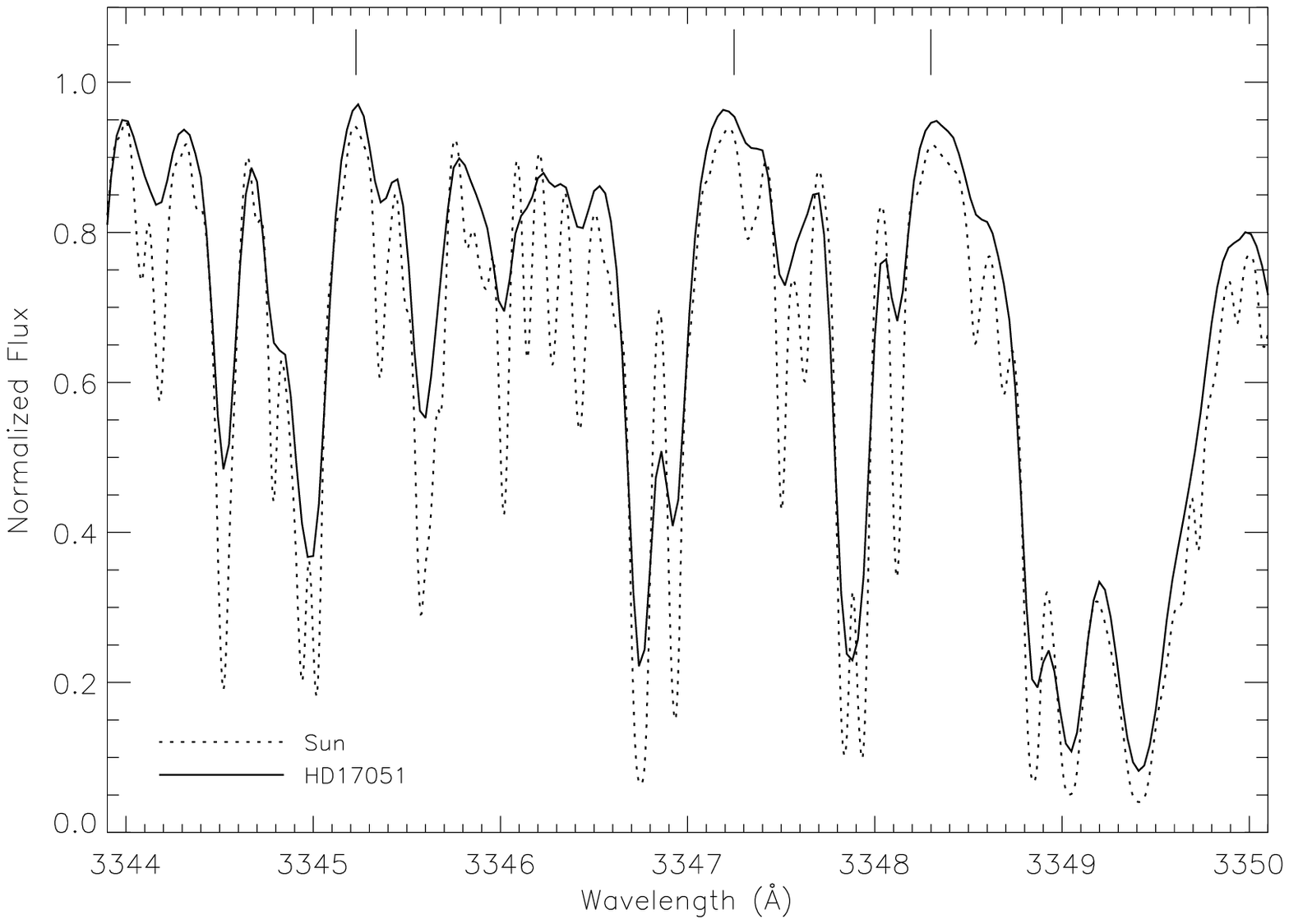}
\includegraphics[height=6cm]{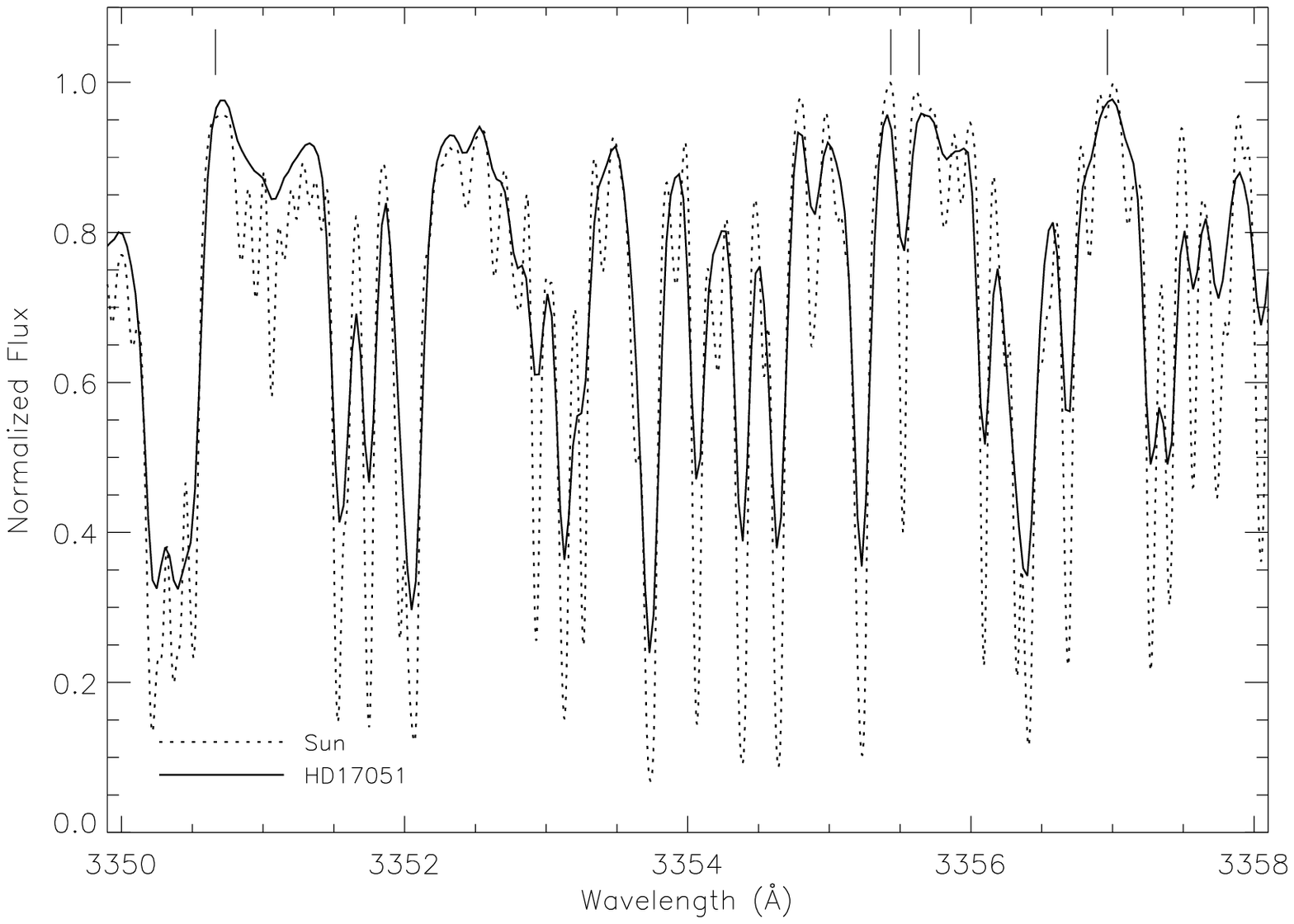}
\centering
\includegraphics[height=6cm]{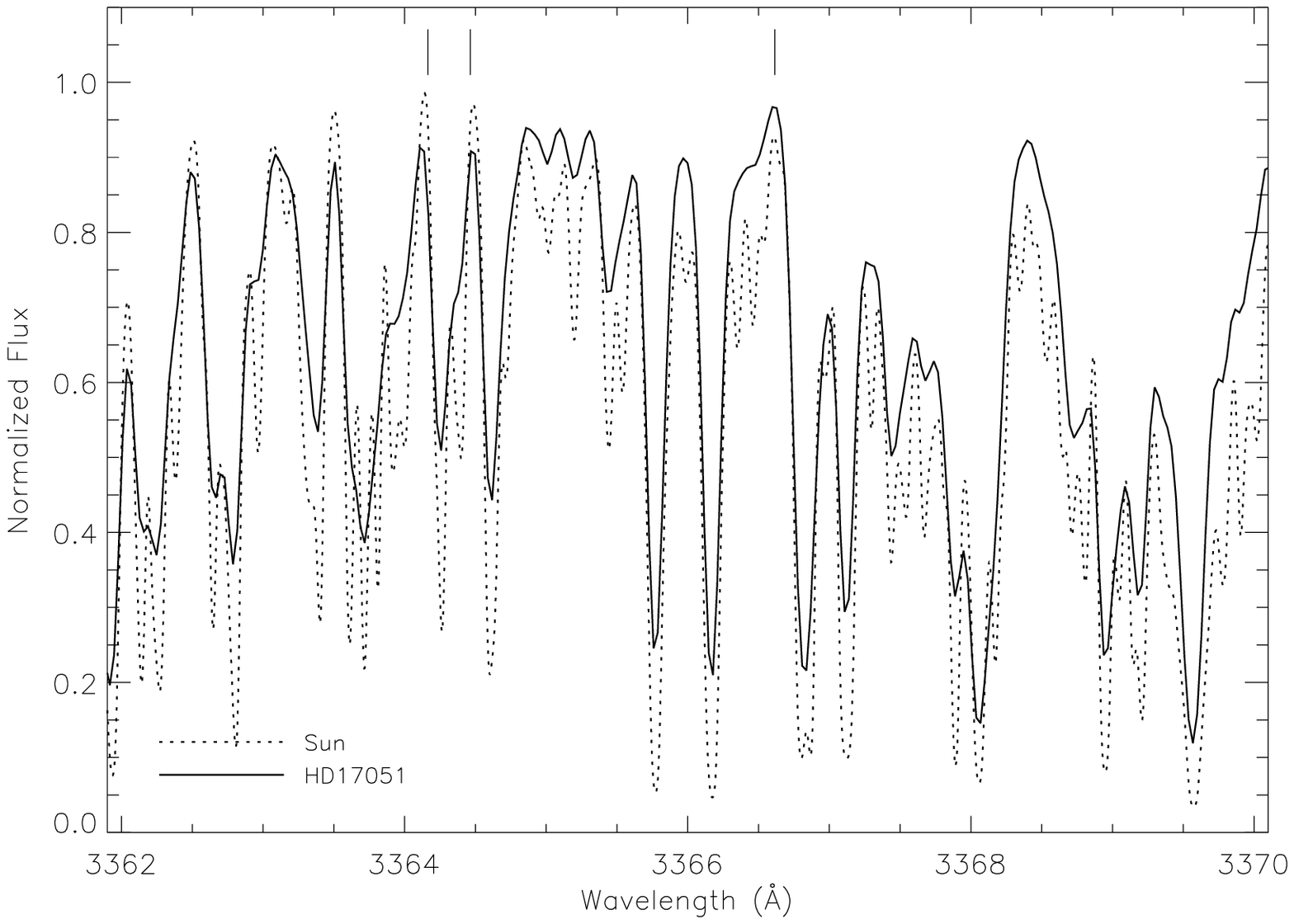}
\caption[]{\object{HD\,17051} and solar spectra in the regions $\lambda\lambda$3344--3350 $\AA$, $\lambda\lambda$3350--3358 
$\AA$ and $\lambda\lambda$3362--3370 $\AA$. 
Vertical lines represent points of reference in continuum determination.}   
\label{fig0}
\end{figure*}

\begin{figure*}
\includegraphics[height=6cm]{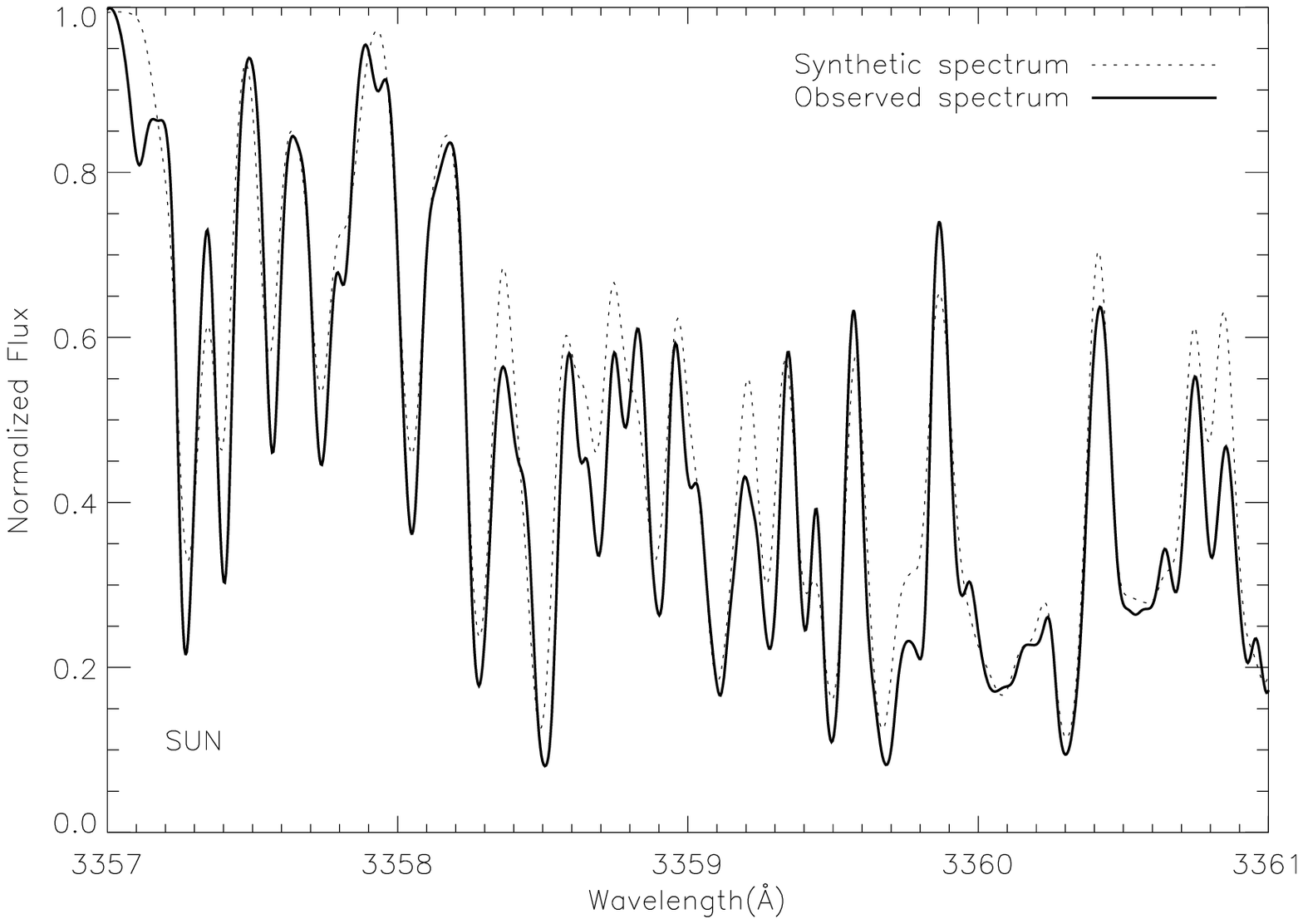}
\includegraphics[height=6cm]{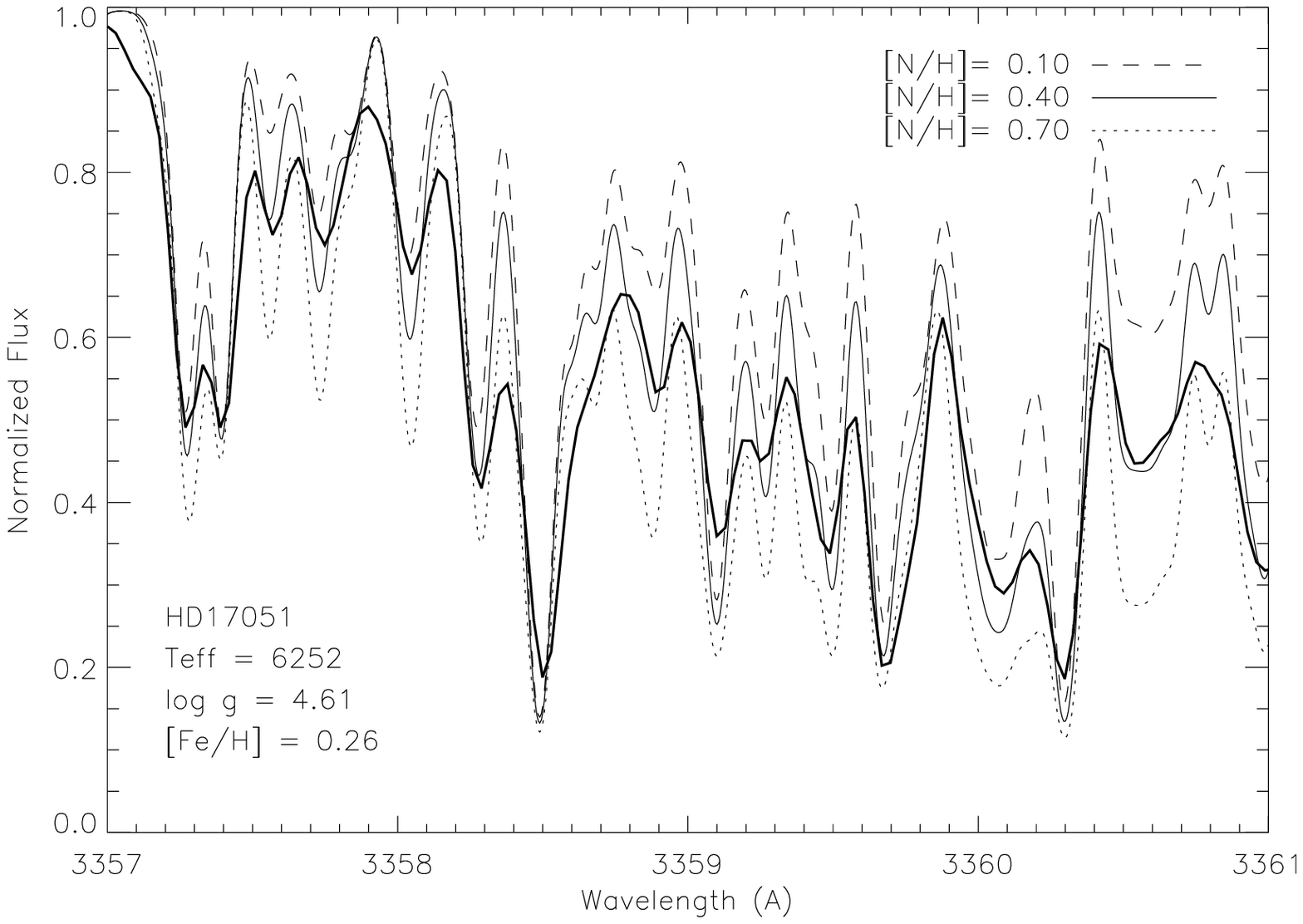}
\includegraphics[height=6cm]{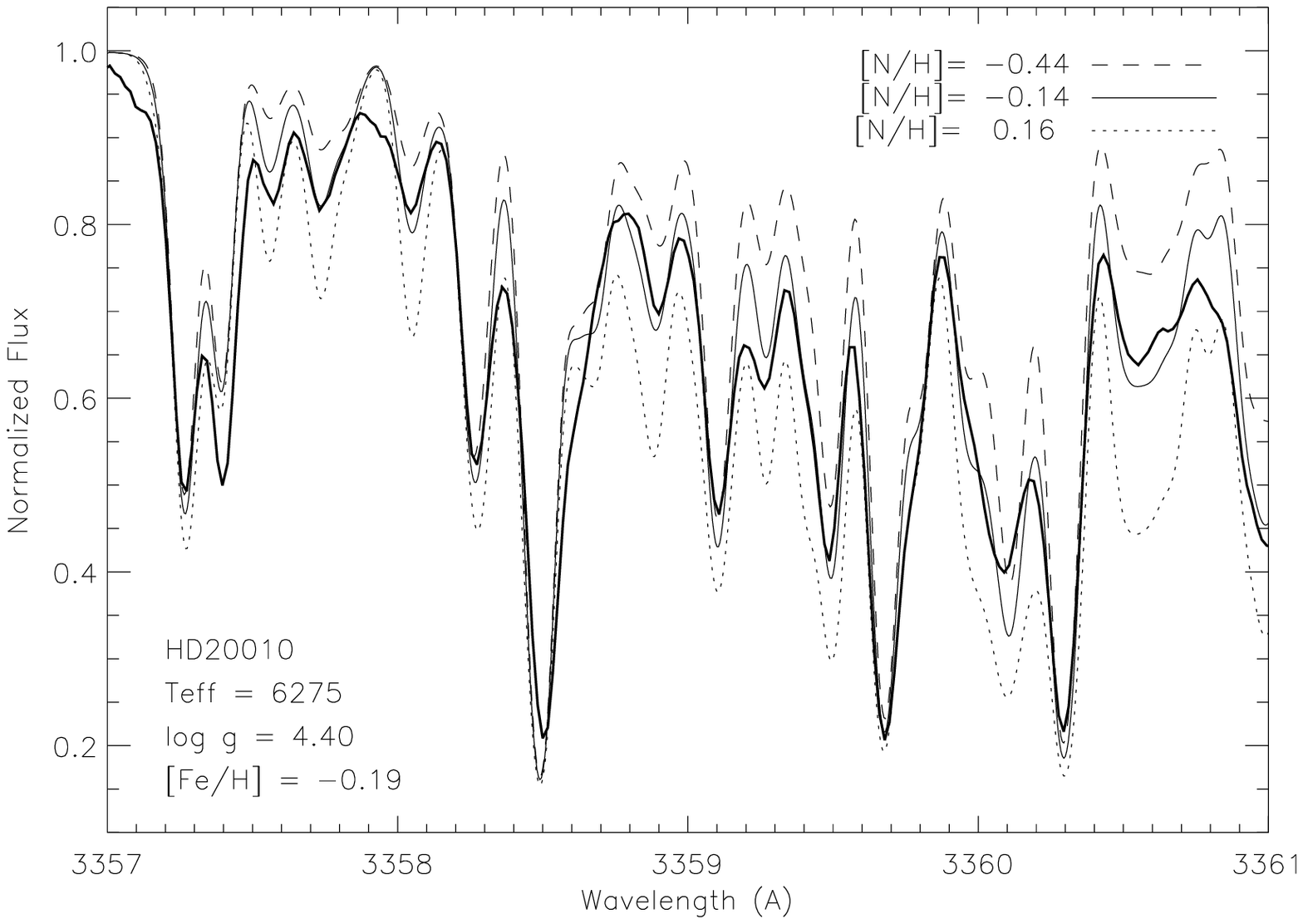}
\includegraphics[height=6cm]{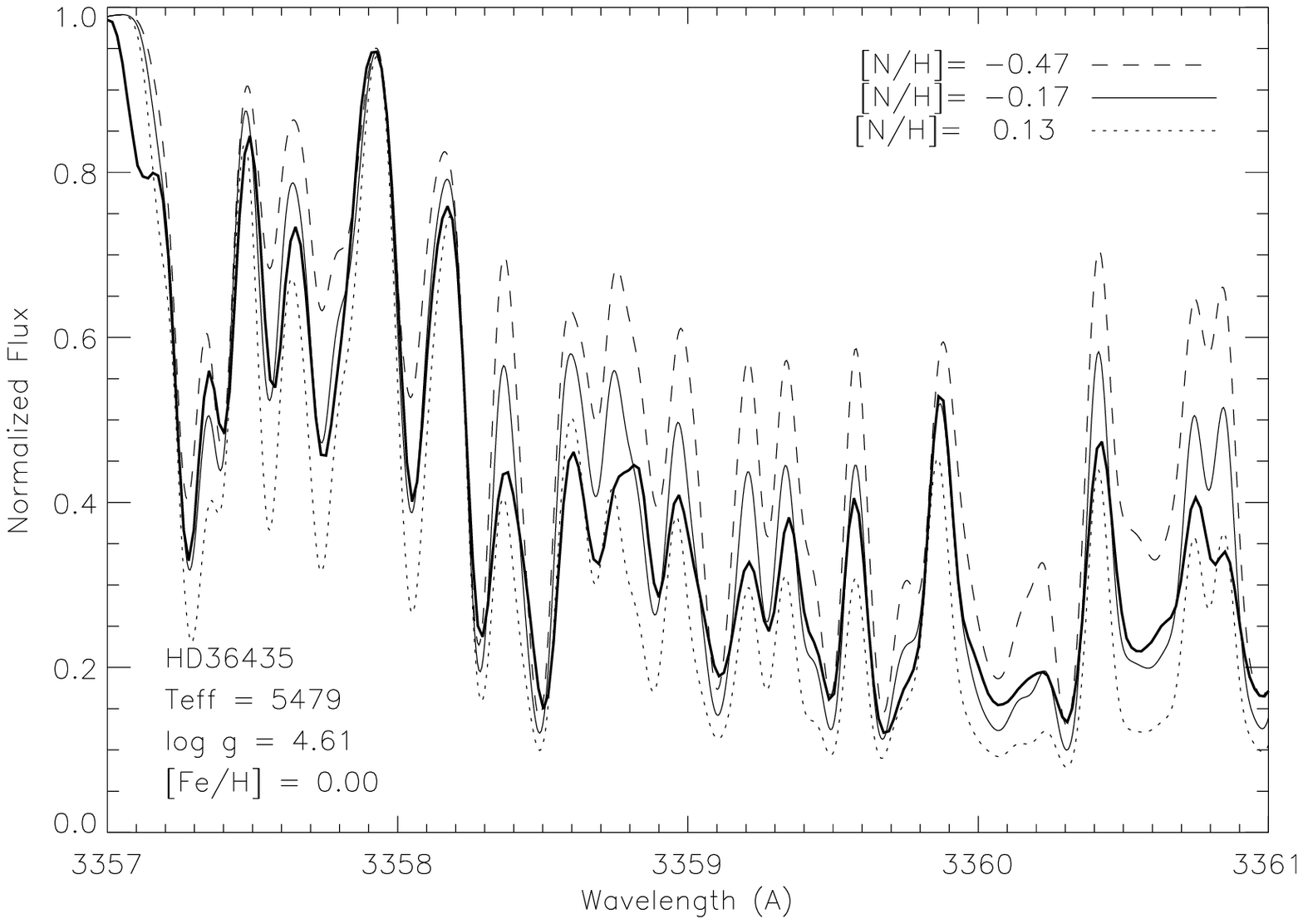}
\caption[]{ Top left: solar observed (solid line) and synthetic (dotted line) spectra in the spectral 
region $\lambda\lambda$3357--3361 \AA. Top right and bottom: observed spectrum (thick solid line) and three 
synthetic spectra (dotted, dashed and solid lines) for different values of [N/H], for three targets.}
\label{fig1}
\end{figure*}

Knowledge of CNO abundances in main sequence stars is clearly important because of the key roles 
of the former in the chain of 
nucleosynthesis. They can give us information about the history of element production in different 
types of stars. 
Solar-type stars can be very informative as tracers of the chemical and dynamical evolution of
the Galaxy (e.g. Pagel \& Edmunds \cite{Pag81}; Edvardsson et al.\ \cite{Edv93}; Bensby et al.\ \cite{Ben03}) as they spend  
between $10^9$ to several times $10^{10}$ years on the main sequence. 

The abundant isotope \element[][14]{N} is synthesized through the CNO cycles in a hydrogen-burning shell of 
 the stellar 
interior. With regard to the chemistry of planet-harbouring stars, the study of the nitrogen abundances of solar-type dwarfs with and
without known planets represents an important key to checking the ``self-enrichment'' scenario. This hypothesis attributes 
the origin of the overabundance of metals to the accretion of large amounts of metal-rich,  H- and He-depleted 
rocky 
planetesimal materials on to the star. 

On the one hand, by studying chemical abundances and planetary orbital parameters, several authors  reject the 
possibility of pollution as the main source of the metallicity enhancement observed in planet host stars and
 propose instead that the metal-rich nature of stars with planets has most probably a ``primordial'' source in the high 
metallicity of the primordial cloud (Santos et al.\ \cite{San01}, \cite{San02}, \cite{San03a}a). 
The effect on metal abundances of contamination in the outer convective zone by accreted planetary material in FGK  
main sequence stars has been discussed in Pinsonneault et al.\ (\cite{Pin01}) and the same conclusion was reached.  
However, the possibility that pollution may occur has not been excluded. Israelian et al.\ (\cite{Isr01}, \cite{Isr03a}) 
have found 
evidence of the infall of a planet into the planet host star \object{HD\,82943}.
\element[][6]{Li} and \element[][7]{Li} may provide an independent test to confirm this scenario 
(Israelian et al.\ \cite{Isr03a}, \cite{Isr03b}).

On the other hand, the ``self-enrichment'' scenario would imply that volatile element (with low condensation 
temperature $T_{\rm C}$)
abundances do not show  overabundance in the same way that refractories do, since the elements of the former group 
are deficient in 
accreted materials relative to the latter. This hypothesis has to be tested by searching for possible differences 
among volatile elements, such as CNO, S and Zn, and refractory ones, such as $\alpha$s Si, Mg, Ca, Ti and the 
iron-group 
elements. 
Smith et al.\ (\cite{Smi01}) found out that a small subset of stars with planets bore this accretion signature. 
 They 
discovered that these stars exhibit a trend of increasing [$X$/H] with increasing $T_{\rm C}$ for a given element $X$. 
However, a following study by Takeda et al.\ (\cite{Tak01}) showed that all the elements, volatiles and refractories alike, 
of fourteen planet-harbouring stars behave quite similar, suggesting that the enhanced metallicity is not caused by the
 accretion of
 rocky 
material but is rather of primordial origin. 
Recently, Sadakane et al.\ (\cite{Sad02}) confirmed this 
result for nineteen elements, volatiles and refractories, in twelve planet host stars. 
Gonzalez et al.\ (\cite{Gonz01}) and Santos et al.\ (\cite{San00}) did not find 
significant differences in the [C/Fe] and [O/Fe] values among planet-harbouring and field stars. Some of these 
authors derived nitrogen abundances from atomic lines (e.g.\ Gonzalez \& Laws \cite{Gonz00}; Gonzalez et al.\ \cite{Gonz01}; 
Takeda et al.\ \cite{Tak01}; Sadakane et al.\ \cite{Sad02}). We note that most of them made no mention of their
final results and conclusions concerning nitrogen.

Various studies have accumulated evidence that the production of
nitrogen at low [Fe/H] proceeds principally as
 a primary 
rather than a secondary process (Pagel \& Edmunds \cite{Pag81}; Bessel \& Norris 
\cite{Bes82}; Carbon et al.\ \cite{Car87}; Henry et al.\ \cite{Hen00}).
The secondary process dominates at high metallicities. 
Two possible primary sources of nitrogen have been proposed. The first is intermediate mass (4--8 $M_{\sun}$) 
stars during 
their thermally pulsing asymptotic giant branch phases: if temperature conditions are suitable, primary nitrogen can be 
produced by CN processing in the convective envelope from freshly synthesized dredged up carbon
(Marigo \cite{Mar01}; van den Hoek \& Groenewegen \cite{van97}). 
The second source is
rotating massive stars (Maeder \& Meynet \cite{Mae00}). The latter would imply uncoupled N and Fe abundances and a nitrogen overproduction relative to 
iron. 

Recently, Liang et al.\ (\cite{Liang01}) and Pettini et al.\ (\cite{Pet02}) 
concluded that the contribution by massive stars to the nitrogen 
abundance is generally very low, 
and
that the dominant N contributors are intermediate and low mass stars (ILMS). Although the
nitrogen is chiefly primary at low [Fe/H], ILMS are
 also 
sources of secondary nitrogen. The relative weights of the secondary and primary components depend on the interplay between 
secondary enrichment caused by dredge-up episodes and the primary contribution given by the CNO cycle during the envelope 
burning.
Shi et al.\ (\cite{Shi02}) obtained similar results by measuring \ion{N}{i} lines in a large sample of disk stars with a
metallicity range of $-1.0<$[Fe/H]$<+$0.2.  

However, 
the situation is rather confused. Other authors support the origin of primary nitrogen in massive stars (e.g.\ 
Izotov \& Thuan \cite{Izot00}) because of the low scatter of [N/O] ratios in galaxies observed at
different stages of their evolution, which would imply no time delay between the injection of nitrogen and oxygen. 
Meynet \& Maeder (\cite{Mey02}) have 
discussed the effect of rotation in the production of primary and secondary nitrogen in intermediate and high mass 
stars,
introducing axial rotation and rotationally induced mixing of chemical species in their stellar models.
They concluded that ILMS are the main source of primary nitrogen at low
metallicities. 

The main problem that has traditionally made it
difficult to measure nitrogen abundances is that there are very few
atomic nitrogen  lines in the red part of the spectrum. Efficient modern detectors now enable us to obtain high-resolution
near-UV spectra, which allows us to carry out a precise and independent study of nitrogen abundance by using 
the NH band at 3360 \AA.

The purpose of this work is to obtain a systematic, uniform and detailed study of
nitrogen abundance in two samples of solar-type dwarfs, a large set of planet hosts and a comparison volume-limited 
sample of stars with no known planetary mass companions.  
We used two independent indicators in the
abundance determination: spectral synthesis of the NH band in near-UV high-resolution spectra and the
near-IR \ion{N}{i} 7468\AA\ line.

\section{Observations}

Spectra for most of the planet-harbouring stars have been collected and used to derive precise stellar parameters in 
a series of recent papers (e.g.\ Santos et al.\ \cite{San01}, \cite{San03a}a, \cite{San03b}b).  
For measurement of the \ion{N}{i} 7468\AA\ line we used all the spectra of this set obtained with the FEROS
spectrograph on the 2.2 m ESO/MPI telescope (La Silla, Chile), the SARG spectrograph on the 3.5 m TNG and
 the 
UES spectrograph on the 4.2 m WHT (both at the Roque de los Muchachos Observatory, La Palma), and other spectra 
from subsequent observational runs using the same instruments. 

For NH band synthesis, we used the same spectra as those in Santos et al.\ (\cite{San02}) to derive
beryllium abundances\footnote{Observing run 66.C-0116A}. New spectra were obtained with UVES spectrograph, at the VLT/UT2 
Kueyen telescope (Paranal Observatory, ESO, Chile)\footnote{Observing run 68.C-0058A}.
The observing log for the spectra from campaigns carried out after Santos et al.\ (\cite{San02}) and
 Bodaghee et al.\ (\cite{Bod03}) are 
listed in Tables~\ref{tab1} and \ref{tab2}.
The obtained high S/N spectra have a minimum resolution $R \ge 50000$.

The data reduction for the SARG and UVES spectra was done using IRAF
tools in the {\tt echelle} package. Standard background correction, flatfield and extraction procedures were used.
The wavelength calibration was performed using a ThAr lamp spectrum taken on the same night. The FEROS spectra were reduced
using the FEROS pipeline software.  

\begin{table*}[t]
\caption[]{Sensitivity of the nitrogen abundance, derived from the 
NH band at 3360 \AA\ and from the \ion{N}{i} line at 7468 \AA, to changes of 100 K in effective temperature, 0.2 dex in
gravity and 0.2 dex in metallicity.}
\begin{center}
\begin{scriptsize}
\begin{tabular}{ccccc}
\hline
\noalign{\smallskip}
 & Star & \object{HD\,46375} & \object{HD\,73526} & \object{HD\, 7570} \\ 
 & ($T_\mathrm{eff}$; $\log {g}$ ; [Fe/H]) & (5268; 4.41; 0.20) & (5699; 4.27; 0.27) & (6140; 4.39; 0.18)\\
\noalign{\smallskip}
\hline 
\noalign{\smallskip}
NH : & $\Delta T_\mathrm{eff}=\pm100$ K & $\pm0.09$ & $\pm0.09$ & $\pm0.09$ \\
\noalign{\smallskip}
\hline 
\noalign{\smallskip}
\ion{N}{i} : & $\Delta T_\mathrm{eff}=\pm100$ K& $\mp0.11$ & $\mp0.08$ & $\mp0.07$ \\
\noalign{\smallskip}
\hline
\noalign{\bigskip}
\hline
\noalign{\smallskip}
 & Star & \object{HD\,19994} & \object{HD\,12661} & \object{HD\, 142415} \\ 
 & ($T_\mathrm{eff}$; $\log {g}$ ; [Fe/H]) & (6190; 4.19; 0.24) & (5702; 4.33; 0.36) & (6045; 4.53; 0.21)\\
\noalign{\smallskip}
\hline 
\noalign{\smallskip}
NH : & $\Delta \log {g}=\pm0.2$ dex & $\mp0.04$ & $\mp0.05$ & $\mp0.05$ \\
\noalign{\smallskip}
\hline 
\noalign{\smallskip}
\ion{N}{i} : & $\Delta \log {g}=\pm0.2$ dex & $\pm0.06$ & $\pm0.06$ & $\pm0.06$ \\
\noalign{\smallskip}
\hline
\noalign{\bigskip}
\hline
\noalign{\smallskip}
 & Star & \object{HD\,143761} & \object{HD\,187123} & \object{HD\, 76700} \\ 
 & ($T_\mathrm{eff}$; $\log {g}$ ; [Fe/H]) & (5853; 4.41; -0.20) & (5845; 4.42; 0.13) & (5737; 4.25; 0.41)\\
\noalign{\smallskip}
\hline 
\noalign{\smallskip}
NH : & $\Delta$([Fe/H]) $= \pm0.2$ dex & $\pm0.14$ & $\pm0.17$ & $\pm0.18$ \\
\noalign{\smallskip}
\hline 
\noalign{\smallskip}
\ion{N}{i} : & $\Delta$([Fe/H]) = $\pm0.2$ dex & $\mp0.03$ & $\mp0.03$ & $\mp0.03$ \\
\noalign{\smallskip}
\hline
\end{tabular}
\end{scriptsize}
\end{center}
\label{tab3}
\end{table*}

\begin{figure*}
\centering
\includegraphics[height=6cm]{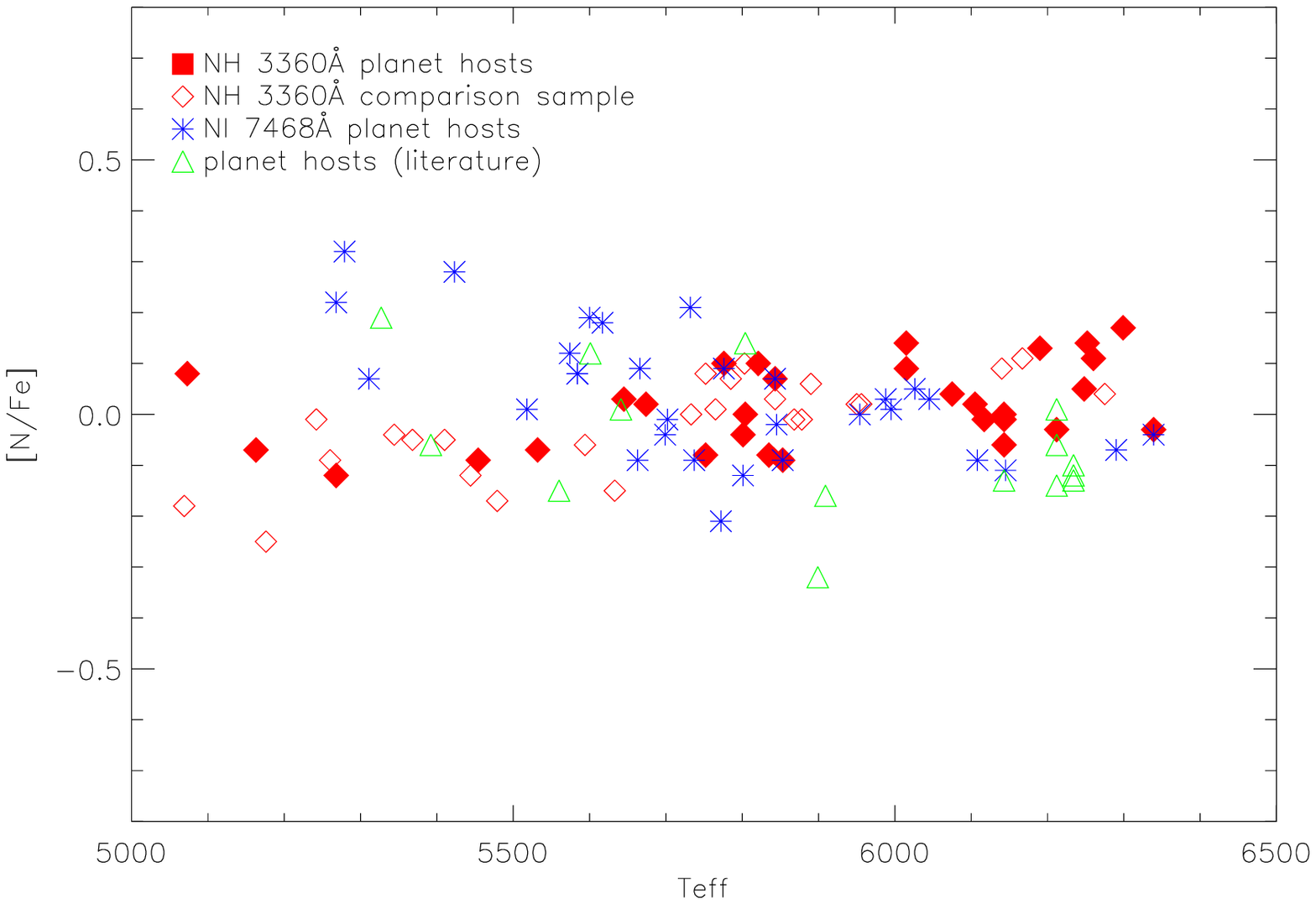}
\includegraphics[height=6cm]{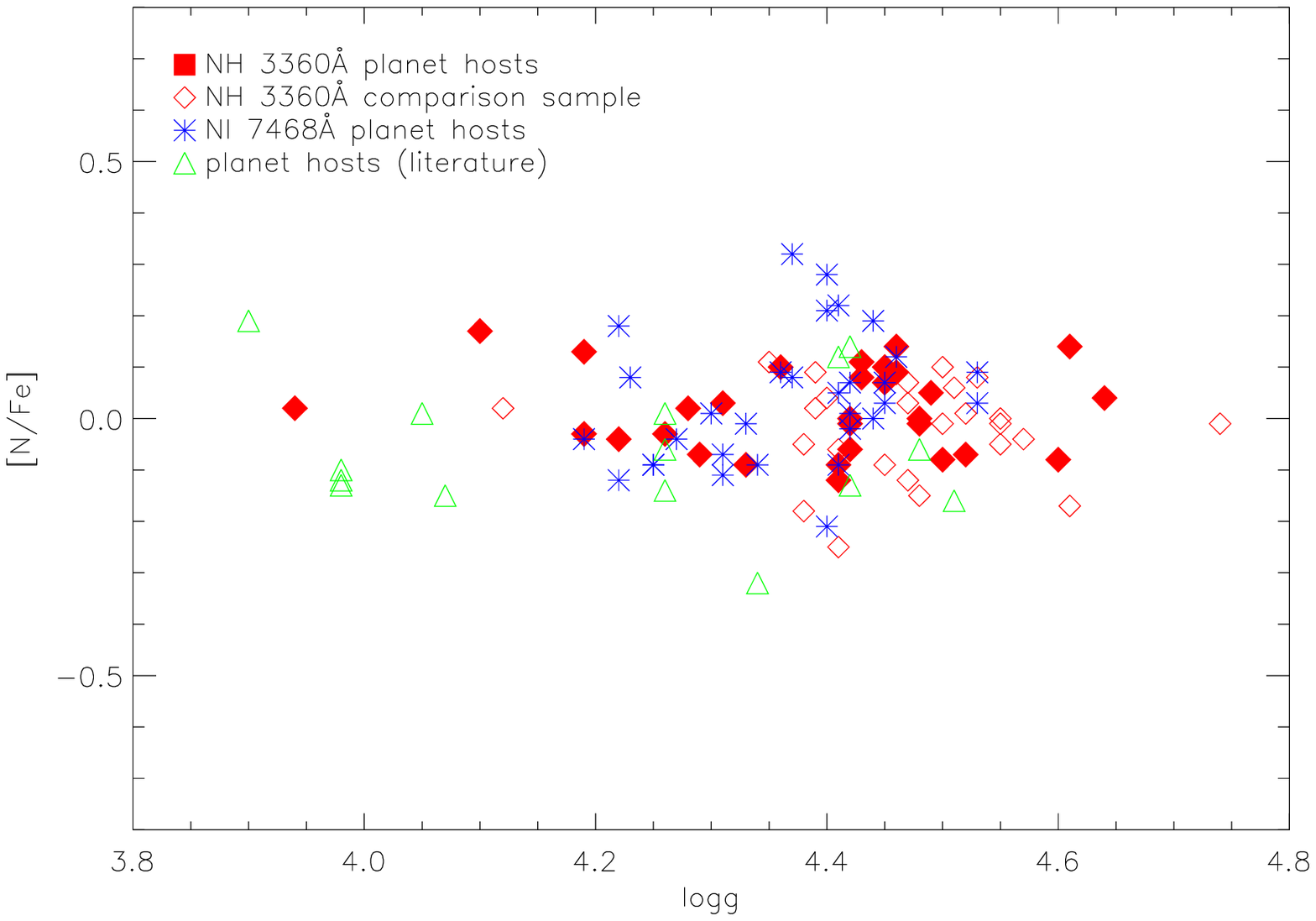}
\caption{[N/Fe] vs.\ $T_\mathrm{eff}$ and $\log {g}$. Filled and open diamonds represent 
planet host and comparison sample stars from NH band synthesis, respectively.
Asterisks and open triangles denote planet host stars from \ion{N}{i} line and from literature
values, respectively.}
\label{fig2}
\end{figure*}

\section{Analysis}

Abundances for all elements were determined using a standard local thermodynamic equilibrium (LTE) analysis 
with the revised
version of the spectral synthesis code MOOG (Sneden \cite{Sne73}) and a grid of Kurucz (\cite{Kur93}) ATLAS9 
atmospheres. All the atmospheric parameters, $T_\mathrm{eff}$, $\log {g}$, $\xi_t$ and [Fe/H], and the corresponding
uncertainties were taken from Santos et al.\ (\cite{San03b}b). The adopted solar abundances for nitrogen and iron 
were $\log{\epsilon}\,(N)_{\odot}$ = 8.05\,dex (Anders \& Grevesse \cite{And89})
and $\log{\epsilon}\,(Fe)_{\odot}$ = 7.47\,dex (Santos et al.\ \cite{San03b}b). 

\subsection{Synthesis of NH band}
\label{synth}

\begin{table*}[t]
\caption[]{Nitrogen abundances from NH band synthesis for a set of stars with planets and brown dwarf companions. Last
column lists $\chi^2_{tot}$ values.}
\begin{center}
\begin{scriptsize}
\begin{tabular}{lcccccr}
\hline
\noalign{\smallskip}
Star & $T_\mathrm{eff}$ & $\log {g}$ & $\xi_t$ & [Fe/H] & [N/H] & $\chi^2_{tot}$ \\
 & (K) & (cm\,s$^{-2}$) & (km\,s$^{-1}$) & & & \\
\hline
\hline
\noalign{\smallskip}
\object{HD\,\,6434} & $5835\pm50$& $4.60\pm0.15$& $1.53\pm0.10$& $-0.52\pm0.05$ & $-0.60\pm0.13$ & 10.13  \\  
\object{HD\,9826}   & $6212\pm64$& $4.26\pm0.13$& $1.69\pm0.16$& $0.13\pm0.08$  & $0.10\pm0.14$  &  6.27 \\
\object{HD\,13445}  & $5163\pm37$& $4.52\pm0.13$& $0.72\pm0.06$& $-0.24\pm0.05$ & $-0.31\pm0.10$ &  7.27 \\
\object{HD\,16141}  & $5801\pm30$& $4.22\pm0.12$& $1.34\pm0.04$& $ 0.15\pm0.04$ &  $0.11\pm0.11$ &  5.16 \\
\object{HD\,17051}  & $6252\pm53$& $4.61\pm0.16$& $1.18\pm0.10$& $ 0.26\pm0.06$ &  $0.40\pm0.13$ &  2.95 \\
\object{HD\,19994}  & $6190\pm57$& $4.19\pm0.13$& $1.54\pm0.13$& $ 0.24\pm0.07$ &  $0.37\pm0.10$ &  5.27 \\
\object{HD\,22049}  & $5073\pm42$& $4.43\pm0.08$& $1.05\pm0.06$& $-0.13\pm0.04$& $-0.05\pm0.12$  & 27.88 \\
\object{HD\,38529}  & $5674\pm40$& $3.94\pm0.12$& $1.38\pm0.05$& $ 0.40\pm0.06$&  $0.42\pm0.12$  & 13.43 \\
\object{HD\,46375}  & $5268\pm55$& $4.41\pm0.16$& $0.97\pm0.06$& $ 0.20\pm0.06$&  $0.07\pm0.13$  & 19.76 \\
\object{HD\,52265}  & $6103\pm52$& $4.28\pm0.15$& $1.36\pm0.09$& $ 0.23\pm0.07$&  $0.25\pm0.10$  &  3.10 \\
\object{HD\,75289}  & $6143\pm53$& $4.42\pm0.13$& $1.53\pm0.09$& $ 0.28\pm0.07$&  $0.21\pm0.13$  &  3.43 \\
\object{HD\,82943}  & $6016\pm30$& $4.46\pm0.08$& $1.13\pm0.04$& $ 0.30\pm0.04$&  $0.39\pm0.10$  &  6.68 \\
\object{HD\,83443}  & $5454\pm61$& $4.33\pm0.17$& $1.08\pm0.08$& $ 0.35\pm0.08$&  $0.26\pm0.14$  & 21.82 \\
\object{HD\,92788}  & $5821\pm41$& $4.45\pm0.06$& $1.16\pm0.05$& $ 0.32\pm0.05$&  $0.42\pm0.12$  & 12.27 \\
\object{HD\,10647}  & $6143\pm31$& $4.48\pm0.08$& $1.40\pm0.08$& $ -0.03\pm0.04$&  $-0.03\pm0.11$&  3.43 \\
\object{HD\,108147} & $6248\pm42$& $4.49\pm0.16$& $1.35\pm0.08$& $ 0.20\pm0.05$&  $0.25\pm0.12$  &  5.58 \\
\object{HD\,120136} & $6339\pm73$& $4.19\pm0.10$& $1.70\pm0.16$& $ 0.23\pm0.07$&  $0.20\pm0.14$  &  1.87 \\
\object{HD\,121504} & $6075\pm40$& $4.64\pm0.12$& $1.31\pm0.07$& $ 0.16\pm0.05$&  $0.20\pm0.12$  &  7.71 \\
\object{HD\,134987} & $5776\pm29$& $4.36\pm0.07$& $1.09\pm0.04$& $ 0.30\pm0.04$&  $0.40\pm0.11$  & 11.96 \\
\object{HD\,143761} & $5853\pm25$& $4.41\pm0.15$& $1.35\pm0.07$& $ -0.21\pm0.04$ & $-0.30\pm0.11$& 12.88  \\
\object{HD\,169830} & $6299\pm41$& $4.10\pm0.02$& $1.42\pm0.09$& $ 0.21\pm0.05$&  $0.38\pm0.12$  &  7.31 \\
\object{HD\,179949} & $6260\pm43$& $4.43\pm0.05$& $1.41\pm0.09$& $ 0.22\pm0.05$&  $0.33\pm0.12$  &  5.72 \\
\object{HD\,202206} & $5752\pm53$& $4.50\pm0.09$& $1.01\pm0.06$& $ 0.35\pm0.06$&  $0.27\pm0.13$  & 14.31 \\
\object{HD\,209458} & $6117\pm26$& $4.48\pm0.08$& $1.40\pm0.06$& $ 0.02\pm0.03$&  $0.01\pm0.11$  &  5.66 \\
\object{HD\,210277} & $5532\pm28$& $4.29\pm0.09$& $1.04\pm0.03$& $ 0.19\pm0.04$&  $0.12\pm0.10$  &  9.14 \\
\object{HD\,217014} & $5804\pm36$& $4.42\pm0.07$& $1.20\pm0.05$& $ 0.20\pm0.05$&  $0.20\pm0.11$  & 10.19 \\
\object{HD\,217107} & $5646\pm34$& $4.31\pm0.10$& $1.06\pm0.04$& $ 0.37\pm0.05$&  $0.40\pm0.10$  & 10.99 \\
\object{HD\,222582 }& $5843\pm38$& $4.45\pm0.07$& $1.03\pm0.06$& $ 0.05\pm0.05$&  $0.12\pm0.11$  & 10.25 \\
\noalign{\smallskip}
\hline
\end{tabular}
\end{scriptsize}
\end{center}
\label{tab4}
\end{table*}

The NH molecular feature is the strongest one in the spectral region $\lambda\lambda$3345--3375 \AA.
We determined nitrogen abundances by fitting synthetic spectra to data in this wavelength range. 
The NH spectra were calculated with the dissociation potential $D_0(NH)=3.37\pm0.06\,eV$ recommended in Grevesse et al.\
(\cite{Gre90}). 

A detailed line list from Yakovina \& Pavlenko (\cite{Yak98}) was adopted. 
These authors obtained a good fit to the Kurucz Solar Atlas (Kurucz et al.\ \cite{Kur84}) by changing oscillator 
strength 
values of the strongest lines and continuum level in a list of atomic and molecular lines from Kurucz's 1993 database.
In our analysis, we slightly modified $\log$ {\it gf} values of this line list\footnote{The full line list is available in 
www.iac.es} (changes smaller than 0.01 dex in most cases) 
in order to avoid modifying the continuum level when fitting the high-resolution Kurucz Solar Atlas (Kurucz et al.\ 
\cite{Kur84}) with a solar model having $T_\mathrm{eff}= 5777$ K, $\log {g}= 4.44$ dex and $\xi_t=$1.0\,km\,s$^{-1}$, 
assuming the solar N abundance of $\log{\epsilon}\,(N)_{\odot}$ = 8.05\,dex.  

All our targets are solar-type dwarfs with 
[Fe/H] values between $-0.6$ and $+0.4$ dex and $T_\mathrm{eff}$ between 5000 and $6300$ K. Stars with lower  
$T_\mathrm{eff}$ were rejected since they would have 
introduced great uncertainties in the abundance results. The discarded targets were \object{HD\,27442}, 
\object{HD\,192263}, \object{HD\,74576} and \object{HD\,23484}, with $T_\mathrm{eff}$ of 4825 K, 4947 K, 5000 K and 5176 K, 
respectively.

The main difficulties were the severe atomic line contamination of the NH features and the crowding of the lines in the
UV region. The oscillator strengths of these lines are not known with high accuracy.
The crowding of blended lines worsens with the increase of metallicity. This characteristic involves difficulties when 
fixing the continuum level since the whole spectral region is depressed by line absorption. 
Continuum was normalized with 5$^{th}$ order polynomia using the CONT task of IRAF.
However we could make further improvements in the placement of continuum using the Kurucz Solar Flux Atlas 
(Kurucz et al.\  \cite{Kur84}) and DIPSO task of the STARLINK software: continuum level was determined by superimposing 
observed and synthetic solar spectra and
selecting some points of reference. Figure~\ref{fig0} shows significant points in continuum determination, selected in the
spectral regions $\lambda\lambda$3344--3350\,\AA\,, $\lambda\lambda$3350--3358\,\AA\, and $\lambda\lambda$3362--3370\,\AA\ 
for \object{HD\,17051}. 
This method was valid for stars with  
$T_\mathrm{eff}$ and metallicities around solar values, while in cooler objects we were not able to assure 
that continuum level could be scaled to the Sun.
  
In order to find the best synthetic spectrum for each star, we used the program FITTING. 
This program allows us to create a grid of
synthetic spectra for several free parameters, such as abundances, atmospheric parameters, metallicity, etc., and to
compare it with the observed spectrum. To generate synthetic spectra, it calls the spectral synthesis code MOOG. The best 
fit is determined by applying a $\chi^2$ minimization method to the spectral regions considered to be the most
significant. 

We used a Gaussian function with a FWHM of 0.07 \AA\ for the instrumental broadening, and a rotational 
broadening function with $vsini$= 4\,km\,s$^{-1}$. No macroturbulence
broadening was used. We assumed the same rotational velocity for all the targets since small differences would 
affect only $\chi^2$ values, not final nitrogen abundances. Only for \object{HD\,120136} and \object{HD\,19994} 
we took $vsini$= 14.5\,km\,s$^{-1}$ and $vsini$= 8.2\,km\,s$^{-1}$ from the CORALIE database. We stress that planet 
searches based on the Doppler technique are biased in favour of old stars with
low $vsini$ and therefore all our targets are slow rotators. In any case, precise values of $vsini$ are not required for 
the abundance analysis since it is well known that
the rotational convolution is not affecting the EW of the line (Gray \cite{Gra92}). Our test with $\chi^2$ analysis confirms
this assumption for the stars considered in our study. In the estimate of damping constants, the Unsold aproximation 
was adopted for all atomic and molecular lines.
We have verified that the damping effects do not affect our final abundances.
   
At first, the abundances of all relevant elements were fitted for a set of stars 
with different atmospheric parameters. Next, the same fits were performed by scaling the abundance of elements other than 
nitrogen to the [Fe/H] value and by changing only the nitrogen abundance. Same results were obtained in both cases, therefore 
all the  following fittings were carried out by scaling elements other than nitrogen to iron. 

The program FITTING introduced the following data in the code MOOG: the atmospheric model, the line list and the range of
 nitrogen abundance of the grid of synthetic spectra. Then $\chi^2$ values were calculated by comparing each 
 synthetic spectrum with the observed one in the following spectral regions: $\lambda\lambda$3344.0--3344.7\,\AA, 
$\lambda\lambda$3345.9--3346.9\,\AA, $\lambda\lambda$3347.2--3348.2\,\AA, $\lambda\lambda$3353.5--3354.5\,\AA, 
$\lambda\lambda$3357.0--3358.0\,\AA, $\lambda\lambda$3358.5--3359.0\,\AA, $\lambda\lambda$3361.8--3362.5\,\AA, 
$\lambda\lambda$3363.0--3363.75\,\AA, $\lambda\lambda$3363.8--3365.5\,\AA, $\lambda\lambda$3368.0--3368.8\,\AA,
$\lambda\lambda$3369.1--3369.9\,\AA, $\lambda\lambda$3371.8--3372.8\,\AA\,and $\lambda\lambda$3374.8--3376.0\,\AA.
$\chi^2_{tot}$ was the sum of $\chi^2$ values\footnote{$ \chi^2=\sum_{i=1}^N (F_i-S_i)^2 $, where $F_i$ and $S_i$ are the observed and 
synthetic spectra fluxes, respectively, at the wavelength i.} of the mentioned spectral regions. 
The nitrogen abundance corresponding to minimum $\chi^2_{tot}$ value was extracted and listed in Tables~\ref{tab4} 
and~\ref{tab5}. The fact that $\chi^2_{tot}$ is smaller for high $T_\mathrm{eff}$ can be understood as the
consecuence of a suitable placement of continuum in stars more similar to the Sun. 

Figure~\ref{fig1} shows fits for a short, significant spectral region. The observed solar spectrum from the Kurucz Solar 
Atlas (Kurucz et al.\ \cite{Kur84}) and synthetic spectrum are represented in the left panel. The case of the planet host
star \object{HD\,17051} is represented in the right panel (the observed spectrum and three synthetic spectra for different 
values of nitrogen abundance).   

Uncertainties in the
 atmospheric parameters are of the order of 50 K in $T_\mathrm{eff}$, 0.12 dex in $\log{g}$, 0.08\,km\,s$^{-1}$ 
in the microturbulence and 0.05 dex in the metallicity (see Santos et al.\ \cite{San03b}b).
In order to know how variations in the atmospheric parameters affect NH abundances, for each parameter a set of three stars 
 having very different parameter values was selected. For each set of stars we then tested [N/H] sensitivity to changes 
in the parameter ($\pm 100$ K in the case of $T_\mathrm{eff}$, $\pm 0.2$ dex in $\log{g}$ and [Fe/H], and 
0.3 km\,s$^{-1}$ in $\xi_t$). The results are shown in Table~\ref{tab3}. No microturbolence effects were included, 
since an increase of 0.3 km\,s$^{-1}$ produced an average decrease of 0.003 dex in nitrogen
abundances, which is negligible with regard to the effects of other parameters. 
Sensitivity of NH abundance was applied in the propagation of each parameter error on 
abundances. Also, a continuum uncertainty effect of the order of 0.1 dex was considered. All effects were added 
quadratically 
to obtain final the uncertainties for the nitrogen abundances. 

\begin{table*}[t]
\caption[]{Nitrogen abundances from NH band synthesis for a set of comparison stars (without giant planets). Last
column lists $\chi^2_{tot}$ values.}
\begin{center}
\begin{scriptsize}
\begin{tabular}{lcccccr}
\hline
\noalign{\smallskip}
Star & $T_\mathrm{eff}$ & $\log {g}$ & $\xi_t$ & [Fe/H] & [N/H] &  $\chi^2_{tot}$ \\
 & (K) & (cm\,s$^{-2}$) & (km\,s$^{-1}$) & & & \\
\hline 
\hline
\noalign{\smallskip}
\object{HD\,1461} & $5785\pm50$ & $4.47\pm0.15$ & $1.00\pm0.10$ & $0.18\pm0.05$ & $0.25\pm0.13$    &   6.72  \\ 
\object{HD\,1581} & $5956\pm44$ & $4.39\pm0.13$ & $1.07\pm0.09$ & $-0.14\pm0.05$ & $-0.11\pm0.12$  &   4.15  \\ 
\object{HD\,3823} & $5950\pm50$ & $4.12\pm0.15$ & $1.00\pm0.10$ & $-0.27\pm0.05$ & $-0.25\pm0.13$  &   4.89  \\ 
\object{HD\,4391} & $5878\pm53$ & $4.74\pm0.15$ & $1.13\pm0.10$ & $-0.03\pm0.06$ & $-0.04\pm0.13$  &   8.21  \\ 
\object{HD\,7570}& $6140\pm41$ & $4.39\pm0.16$ & $1.50\pm0.08$ &  $0.18\pm0.05$ &   $0.26\pm0.12$  &   6.28  \\ 
\object{HD\,10700} & $5344\pm29$ & $4.57\pm0.09$ & $0.91\pm0.06$ & $-0.52\pm0.04$ & $-0.56\pm0.11$ &  15.42 \\  
\object{HD\,14412} & $5368\pm24$ & $4.55\pm0.05$ & $0.88\pm0.05$ & $-0.47\pm0.03$ & $-0.53\pm0.11$ &  16.58 \\  
\object{HD\,20010} & $6275\pm57$ & $4.40\pm0.37$ & $2.41\pm0.41$ & $-0.19\pm0.06$ & $-0.14\pm0.15$ &   3.69  \\ 
\object{HD\,20766} & $5733\pm31$ & $4.55\pm0.10$ & $1.09\pm0.06$ & $-0.21\pm0.04$ & $-0.21\pm0.11$ &   8.35 \\  
\object{HD\,20794} & $5444\pm31$ & $4.47\pm0.07$ & $0.98\pm0.06$ & $-0.38\pm0.04$ & $-0.50\pm0.11$ &  14.72 \\  
\object{HD\,20807} & $5843\pm26$ & $4.47\pm0.10$ & $1.17\pm0.06$ & $-0.23\pm0.04$ & $-0.20\pm0.11$ &   7.44 \\  
\object{HD\,23484} & $5176\pm45$ & $4.41\pm0.17$ & $1.03\pm0.06$ &  $0.06\pm0.05$ & $-0.19\pm0.12$ &  16.86 \\  
\object{HD\,30495} & $5868\pm30$ & $4.55\pm0.10$ & $1.24\pm0.05$ &  $0.02\pm0.04$ & $0.01\pm0.11$  &   7.86 \\  
\object{HD\,36435} & $5479\pm37$ & $4.61\pm0.07$ & $1.12\pm0.05$ & $-0.00\pm0.05$ & $-0.17\pm0.11$ &   8.82 \\  
\object{HD\,38858} & $5752\pm32$ & $4.53\pm0.07$ & $1.26\pm0.07$ & $-0.23\pm0.05$ & $-0.15\pm0..11$&  10.46 \\  
\object{HD\,43162} & $5633\pm35$ & $4.48\pm0.07$ & $1.24\pm0.05$ & $-0.01\pm0.04$ & $-0.16\pm0.11$ &   6.93 \\  
\object{HD\,43834} & $5594\pm36$ & $4.41\pm0.09$ & $1.05\pm0.04$ &  $0.10\pm0.05$ &  $0.04\pm0.11$ &  11.53 \\  
\object{HD\,69830} & $5410\pm26$ & $4.38\pm0.07$ & $0.89\pm0.03$ & $-0.03\pm0.04$ & $-0.08\pm0.11$ &  15.02 \\  
\object{HD\,72673} & $5242\pm28$ & $4.50\pm0.09$ & $0.69\pm0.05$ & $-0.37\pm0.04$ & $-0.38\pm0.11$ &  16.37 \\  
\object{HD\,76151} & $5803\pm29$ & $4.50\pm0.08$ & $1.02\pm0.04$ &  $0.14\pm0.04$ &  $0.25\pm0.11$ &  12.54 \\  
\object{HD\,84117} & $6167\pm37$ & $4.35\pm0.10$ & $1.42\pm0.09$ & $-0.03\pm0.05$ &  $0.09\pm0.12$ &   4.16 \\  
\object{HD\,189567} & $5765\pm24$ & $4.52\pm0.05$ & $1.22\pm0.05$ & $-0.23\pm0.04$ & $-0.22\pm0.11$&   9.65 \\  
\object{HD\,192310} & $5069\pm49$ & $4.38\pm0.19$ & $0.79\pm0.07$ & $-0.01\pm0.05$ & $-0.19\pm0.13$&  19.83 \\  
\object{HD\,211415} & $5890\pm30$ & $4.51\pm0.07$ & $1.12\pm0.07$ & $-0.17\pm0.04$ & $-0.11\pm0.11$&   7.88 \\  
\object{HD\,222335} & $5260\pm41$ & $4.45\pm0.11$ & $0.92\pm0.06$ & $-0.16\pm0.05$ & $-0.25\pm0.12$&  12.98 \\  
\noalign{\smallskip}
\hline
\end{tabular}
\end{scriptsize}
\end{center}
\label{tab5}
\end{table*}

\subsection{The \ion{N}{i} absorption at 7468.27 \AA}
\label{NIan}
\begin{figure*}
\centering
\includegraphics[height=6cm]{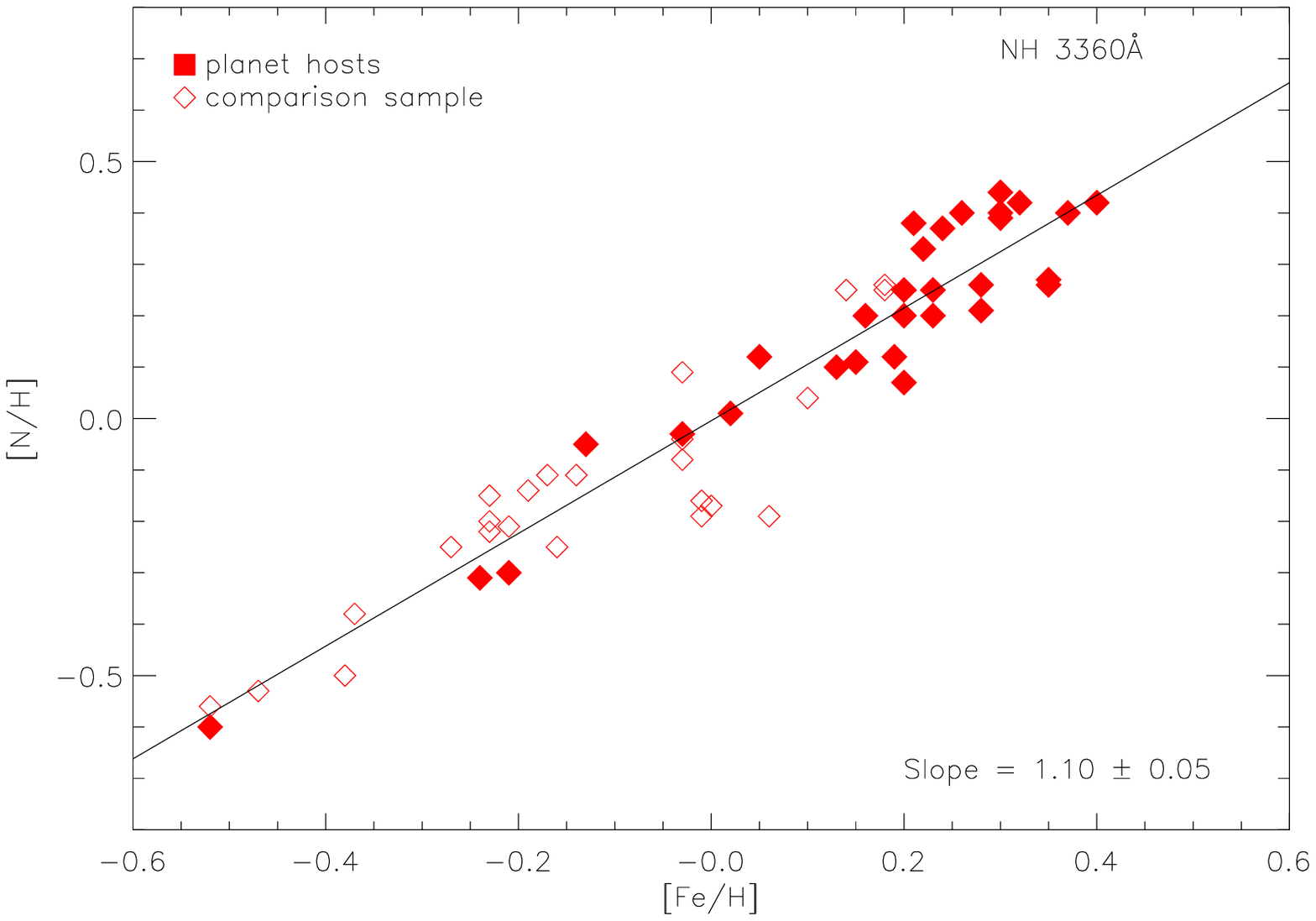}
\includegraphics[height=6cm]{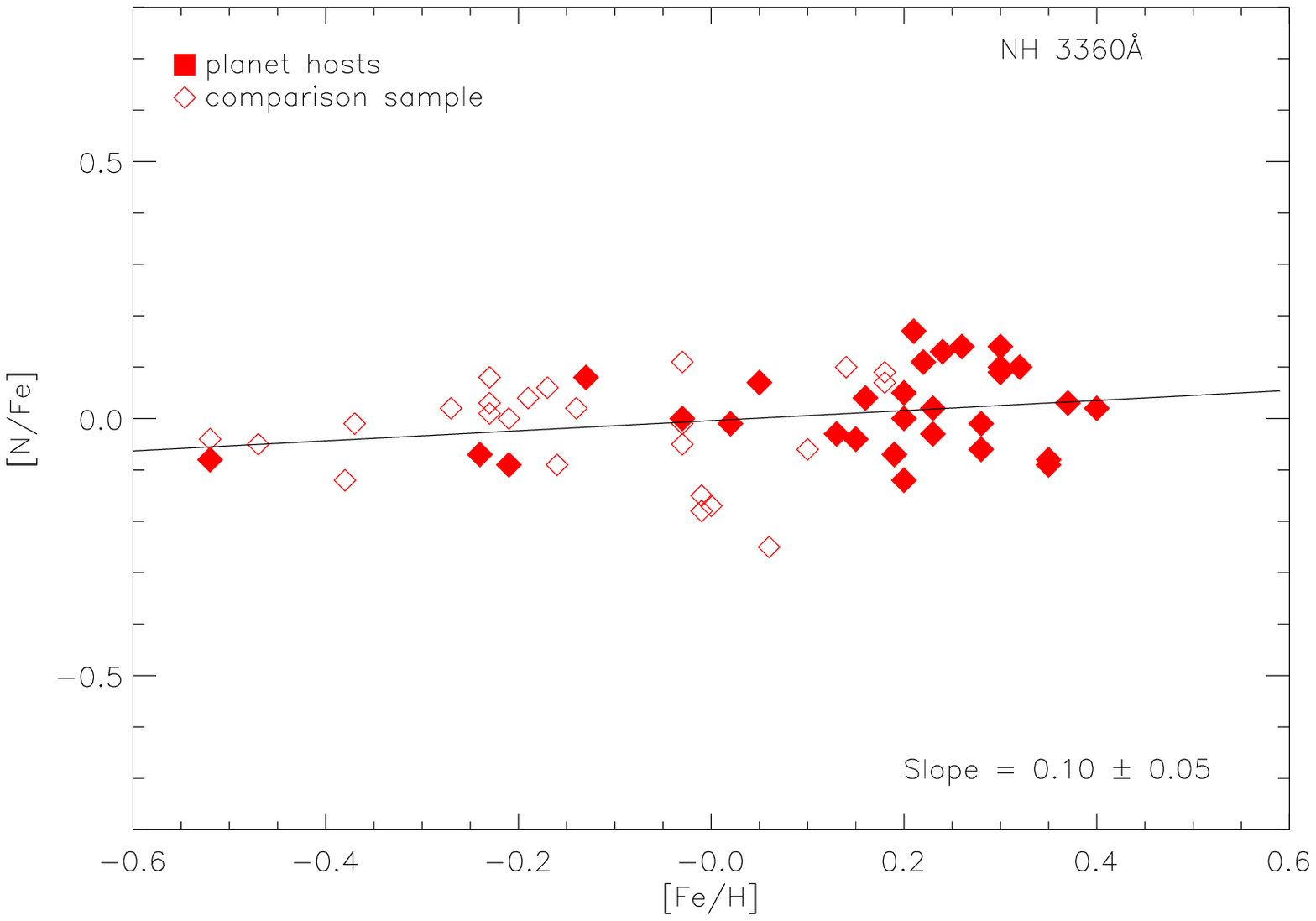}
\caption{[N/H] and [N/Fe] vs.\ [Fe/H] plots from NH 3360 \AA\ band synthesis. Filled and open diamonds represent planet
 host 
and comparison sample stars, respectively. Linear least-squares fits to both samples together are represented and 
slope values are written at the bottom of each plot.}
\label{fig3}
\end{figure*}

\begin{figure*}
\centering
\includegraphics[width=9cm]{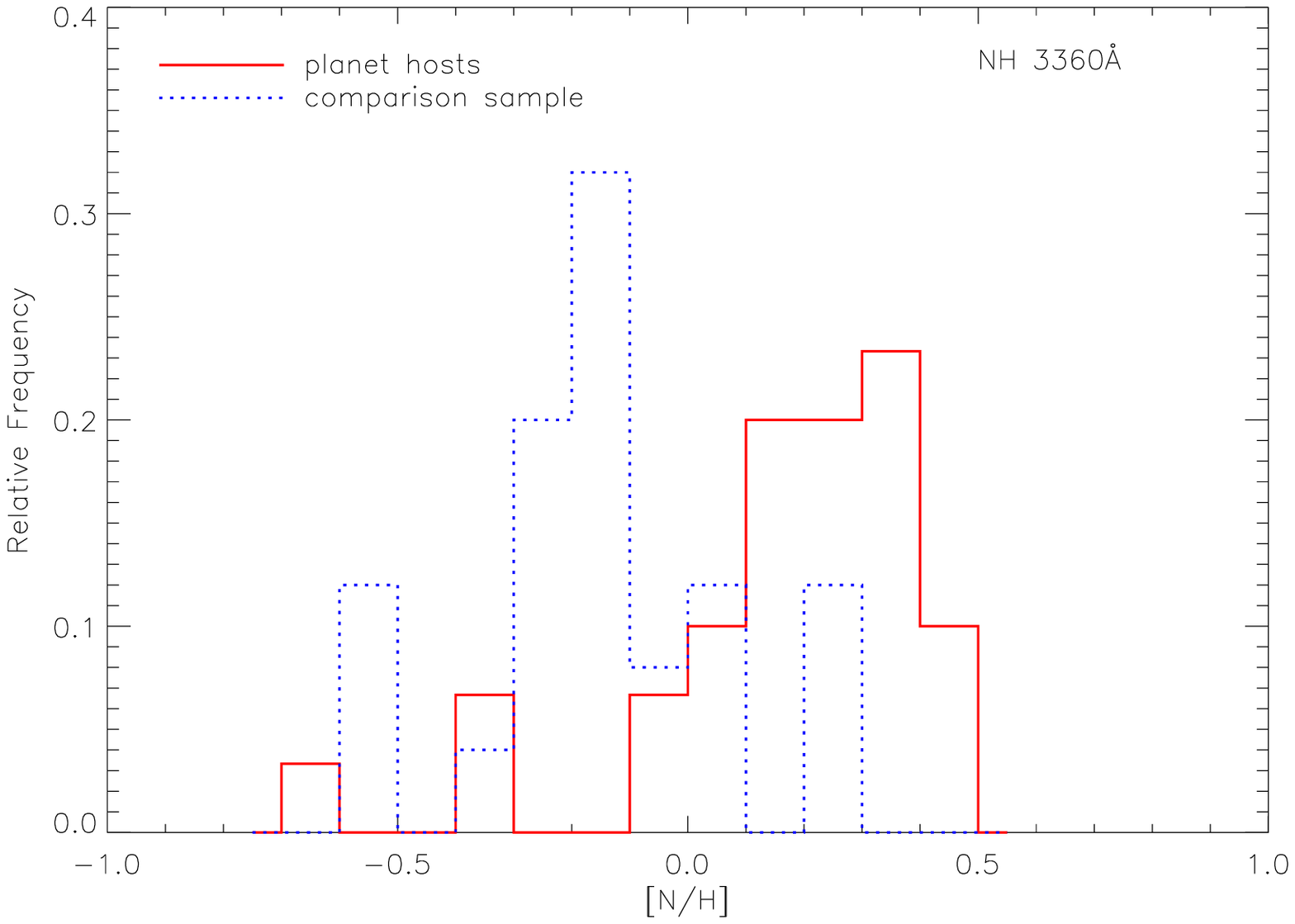}
\includegraphics[width=9cm]{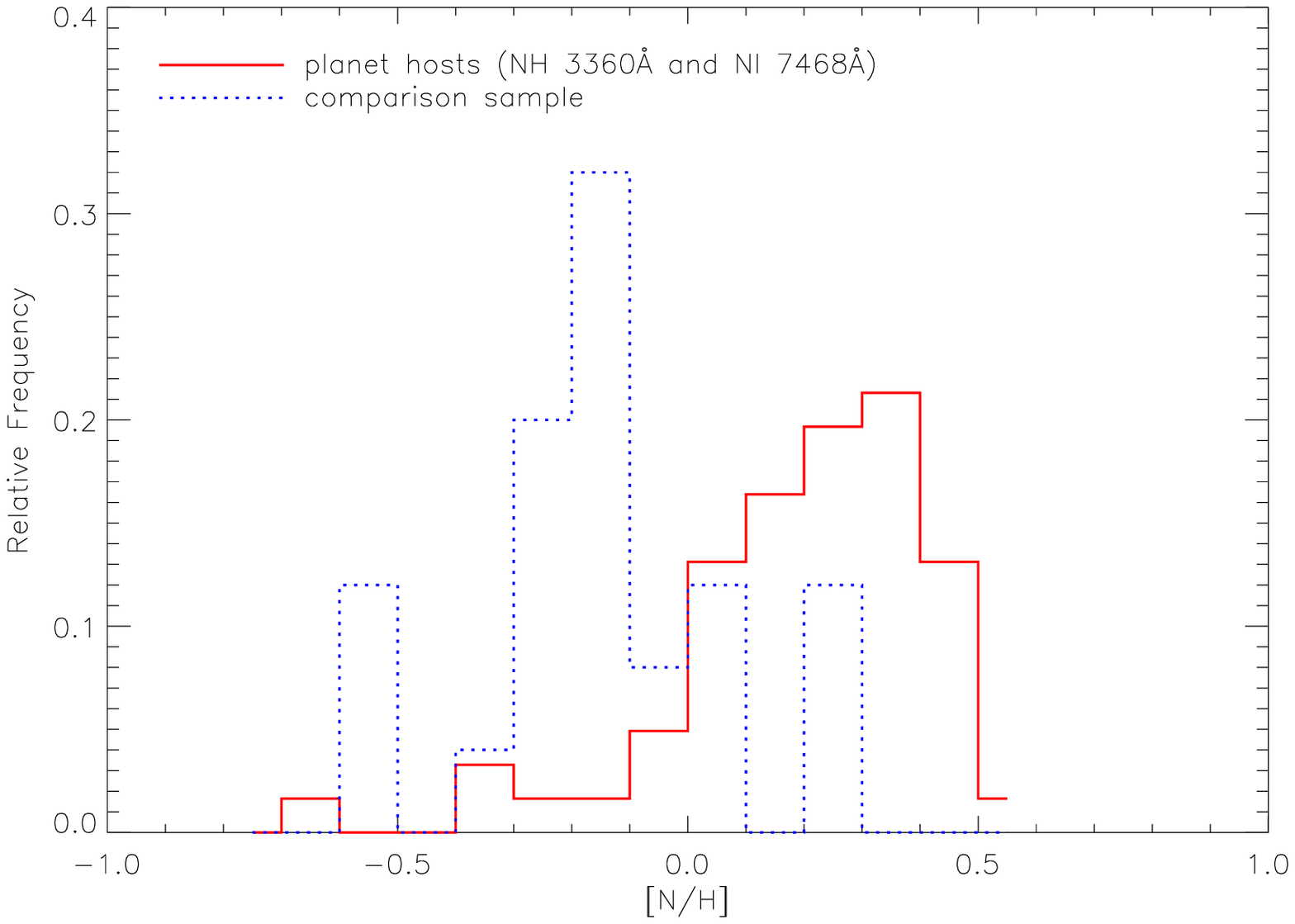}
\includegraphics[width=9cm]{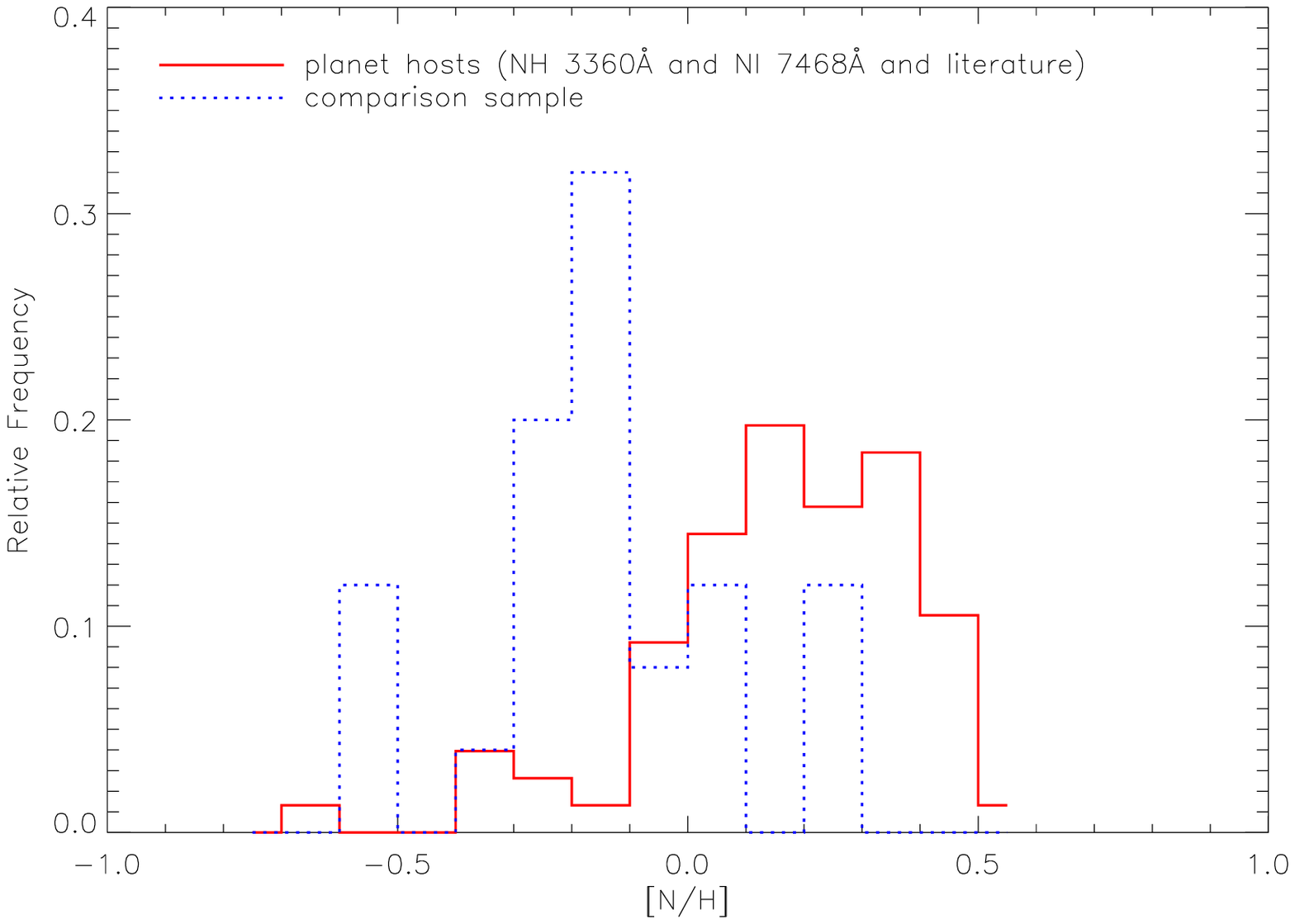}
\caption{[N/H] distributions, from NH band values (upper panel), from NH band and \ion{N}{i} line values (top lower
panel), and from NH band, \ion{N}{i} line and literature values  (bottom panel). Solid and dotted lines represent planet
 host and 
comparison sample stars, respectively.}
\label{fig4}
\end{figure*}

Nitrogen abundance analysis based on the feature at 7468.27 \AA\ was carried out diferentially with respect to the Sun. 
Wavelength and excitation energies of lower level were taken from Moore et al.\ (\cite{Moo66}) and a solar 
gf value ($\log {gf}=-0.15$) was computed using equivalent width measured in the Kurucz Solar Atlas 
(Kurucz et al.\ \cite{Kur84}) (4.2 m\AA) and a solar model with $T_\mathrm{eff}= 5777$ K,
$\log {g}$ = 4.44 dex and $\xi_t=$1.0\,km\,s$^{-1}$, assuming the solar N abundance of 
$\log{\epsilon}\,(N)_{\odot}$ = 8.05\,dex. Equivalent widths in the observed stars were determined by 
Gaussian fitting using the SPLOT task of IRAF, and abundances were computed using the ABFIND driver of MOOG.

The high resolution spectra we used provided us with very reliable measurements in most cases. 
However, a strong fringing effect around 7468 \AA\ prevented us from measuring high 
precision EW values in some cases; therefore, the corresponding EW uncertainties are more significant.

The sensitivity of the \ion{N}{i} line to variations in atmospheric parameters was estimated in the same way as
in the NH case (see Subsection~\ref{synth}). The results are shown in Table~\ref{tab3}.
In order to take into account inaccuracies caused by the continuum determination, equivalent widths for the
highest and the lowest 
continuum levels were measured,  and the corresponding abundance errors were added in quadrature to
the  abundance uncertainties 
derived from inaccuracies in the  atmospheric parameters. 

Dependences on $T_\mathrm{eff}$ and on $\log{g}$ of [N/Fe] performed by both methods are presented in Figure~\ref{fig2}. We 
note that no characteristic trends appear in either case. This means that our results are almost free from systematic 
errors.

\begin{table*}[t]
\caption[]{Nitrogen abundances from \ion{N}{i}
7468.27 \AA\ for a set of stars with planets
and brown dwarf companions.}
\begin{scriptsize}
\begin{center}
\begin{tabular}{lcccccr}
\hline
\noalign{\smallskip}
Star & $T_\mathrm{eff}$ & $\log {g}$       & $\xi_t$   & [Fe/H]  & EW &  [N/H]  \\ 
     & (K)              & \, (cm\,s$^{-2}$) & \, (km\,s$^{-1}$) &     & (m\AA) &  \\
\hline 
\hline
\noalign{\smallskip}
\object{HD\,12661} & $5702\pm36$ & $4.33\pm0.08$& $1.05\pm0.04$& $0.36\pm0.05$& $8.5\pm1$  & $0.35\pm0.07$           \\ 
\object{HD\,16141} & $5801\pm30$ & $4.22\pm0.12$& $1.34\pm0.04$& $0.15\pm0.04$& $5.6\pm1$ & $0.03\pm0.10$            \\ 
\object{HD\,19994} & $6290\pm58$ & $4.31\pm0.13$& $1.63\pm0.12$& $0.24\pm0.07$& $13.0\pm2$ & $0.17\pm0.09$           \\ 
\object{HD\,23596} & $6108\pm36$ & $4.25\pm0.10$& $1.30\pm0.05$& $0.31\pm0.05$& $12.6\pm2$ & $0.22\pm0.10$           \\ 
\object{HD\,30177}	& $5584\pm65$ & $4.23\pm0.13$& $1.14\pm0.07$& $0.38\pm0.09$& $ 9.0\pm2$ & $0.47\pm0.10$      \\ 
\object{HD\,40979} & $6145\pm42$ & $4.31\pm0.15$& $1.29\pm0.09$& $0.21\pm0.05$& $10.0\pm3$ & $0.10\pm0.16$           \\ 
\object{HD\,46375} & $5268\pm55$ & $4.41\pm0.16$& $0.97\pm0.06$& $0.20\pm0.06$& $4.0\pm1$ & $0.42\pm0.15$            \\ 
\object{HD\,50554} & $6026\pm30$ & $4.41\pm0.13$& $1.11\pm0.06$& $0.01\pm0.04$& $7.0\pm1$ & $0.06\pm0.08$            \\ 
\object{HD\,65216} & $5666\pm31$ & $4.53\pm0.09$& $1.06\pm0.05$& $-0.12\pm0.04$ & $3.0\pm1$ & $-0.03^{+0.15}_{-0.19}$\\ 
\object{HD\,68988} & $5988\pm52$ & $4.45\pm0.15$& $1.25\pm0.08$& $0.36\pm0.06$& $13.0\pm3$ & $0.39\pm0.15$           \\ 
\object{HD\,72659}	& $5995\pm45$ & $4.30\pm0.07$& $1.42\pm0.09$& $0.03\pm0.06$& $ 7.0\pm2$ & $0.04\pm0.16$      \\ 
\object{HD\,73256}	& $5518\pm49$ & $4.42\pm0.12$& $1.22\pm0.06$& $0.26\pm0.06$& $ 5.0\pm1$ & $0.27\pm0.11$      \\ 
\object{HD\,73526}	& $5699\pm49$ & $4.27\pm0.12$& $1.26\pm0.06$& $0.27\pm0.06$& $ 7.5\pm2$ & $0.23\pm0.16$      \\ 
\object{HD\,75732} & $5279\pm62$ & $4.37\pm0.18$& $0.98\pm0.07$& $0.33\pm0.07$& $6.5\pm1$ & $0.65\pm0.11$            \\ 
\object{HD\,76700}	& $5737\pm34$ & $4.25\pm0.14$& $1.18\pm0.04$& $0.41\pm0.05$& $ 9.0\pm2$ & $0.32\pm0.14$      \\ 
\object{HD\,80606} & $5574\pm72$ & $4.46\pm0.20$& $1.14\pm0.09$& $0.32\pm0.09$& $7.5\pm2$ & $0.44\pm0.17$            \\ 
\object{HD\,95128} & $5954\pm25$ & $4.44\pm0.10$& $1.30\pm0.04$& $0.06\pm0.03$& $6.4\pm2$ & $0.06\pm0.16$            \\ 
\object{HD\,120136} & $6339\pm73$ & $4.19\pm0.10$& $1.70\pm0.16$& $0.23\pm0.07$& $16.2\pm2$ & $0.19\pm0.09$          \\ 
\object{HD\,134987} & $5776\pm29$ & $4.36\pm0.07$& $1.09\pm0.04$& $0.30\pm0.04$& $10.0\pm2$ & $0.39\pm0.12$          \\ 
\object{HD\,142415} & $6045\pm44$ & $4.53\pm0.08$& $1.12\pm0.07$& $0.21\pm0.05$& $10.0\pm2$ & $0.24\pm0.13$          \\ 
\object{HD\,143761} & $5853\pm25$ & $4.41\pm0.15$& $1.35\pm0.07$&$-0.21\pm0.04$& $2.5\pm1$& $-0.29^{+0.17}_{-0.25}$  \\ 
\object{HD\,145675} & $5311\pm87$ & $4.42\pm0.18$& $0.92\pm0.10$& $0.43\pm0.08$& $5.3\pm2$ & $0.50\pm0.19$           \\ 
\object{HD\,168443} & $5617\pm35$ & $4.22\pm0.05$& $1.21\pm0.05$& $0.06\pm0.05$& $6.0\pm3$ & $0.24^{+0.23}_{-0.36}$  \\ 
\object{HD\,178911B} & $5600\pm42$ & $4.44\pm0.08$& $0.95\pm0.05$& $0.27\pm0.05$& $8.0\pm2$ & $0.46\pm0.17$          \\ 
\object{HD\,186427} & $5772\pm25$ & $4.40\pm0.07$& $1.07\pm0.04$& $0.08\pm0.04$& $3.4\pm1$ & $-0.13\pm0.14$          \\ 
\object{HD\,187123} & $5845\pm22$ & $4.42\pm0.07$& $1.10\pm0.03$& $0.13\pm0.03$& $6.1\pm2$ & $0.11\pm0.16$           \\ 
\object{HD\,190360} & $5584\pm36$ & $4.37\pm0.06$& $1.07\pm0.05$& $0.24\pm0.05$& $6.3\pm2$ & $0.32\pm0.16$           \\ 
\object{HD\,216770} & $5423\pm41$ & $4.40\pm0.13$& $1.01\pm0.05$& $0.26\pm0.04$& $7.0\pm2$ & $0.54\pm0.19$           \\ 
\object{HD\,217107} & $5663\pm36$ & $4.34\pm0.08$& $1.11\pm0.04$& $0.37\pm0.05$& $8.7\pm1$ & $0.28\pm0.06$           \\ 
\object{HD\,219542B} & $5732\pm31$ & $4.40\pm0.05$& $0.99\pm0.04$& $0.17\pm0.04$& $8.0\pm1$ & $0.39\pm0.13$          \\ 
\object{HD\,222582} & $5843\pm38$ & $4.45\pm0.07$& $1.03\pm0.06$& $0.05\pm0.05$& $6.0\pm2$ & $0.12\pm0.16$           \\ 
\noalign{\smallskip}
\hline
\end{tabular}
\end{center}
\end{scriptsize}
\label{tab6}
\end{table*}

\section{Nitrogen abundances from near-UV spectra}
\label{NHtrend}

In the literature there are already a few studies that use NH band at 3360 \AA\ to determine 
nitrogen abundances by spectral synthesis, but only for metal-poor dwarfs (Bessel \& Norris
\cite{Bes82}; Tomkin \& Lambert \cite{Tom84}; Laird \cite{Laird85}; Carbon et al.\ \cite{Car87}).

We analysed near-UV high-resolution spectra of 28 planet host stars and 25 comparison sample stars with the goal of 
checking for possible differences in nitrogen abundances between both samples. 
All the results for planet host and comparison sample stars are presented in Tables~\ref{tab4} and~\ref{tab5},
 respectively.

Figure~\ref{fig3} shows the [N/H] and [N/Fe] abundance ratios as functions of [Fe/H] for these two samples. These
plots indicate that the abundance trend for stars with planets is almost indistinguishable from that of the
comparison stars. Because of the iron enhancement that planet-harbouring stars present, their abundance distributions 
are the 
high [Fe/H] extensions of the curves traced by the comparison sample stars. No peculiar behaviour of the nitrogen abundance 
seems to be associated with the presence of planets.

On the whole, the [N/Fe] vs.\ [Fe/H] trend appears to be almost flat, 
since the slope value is $0.10\pm0.05$. Nevertheless, a small trend
to increase with [Fe/H] cannot be excluded. The [N/H] vs.\ [Fe/H] plot
shows that the behaviour of the N abundance is not significantly different from that of Fe; both elements keep pace
with each other. Moreover, no significant nitrogen overproduction relative to iron is observed. 

The [N/H] distributions for both samples are represented in Fig.~\ref{fig4} (upper panel). 
An interesting feature of the histogram is
that the planet-harbouring stars set does not present a symmetrical distribution, as the comparison sample does. The
 distribution increases for increasing
[N/H] values until reaching the highest value, not centred on the [N/H] range. However, this peculiarity could be caused by
the lack of objects with metallicities higher than $0.5$ dex. Santos et al.\ (\cite{San01}, \cite{San03a}a, \cite{San03b}b) 
found a similar shape in iron distribution.

\begin{figure*}
\centering
\includegraphics[height=6cm]{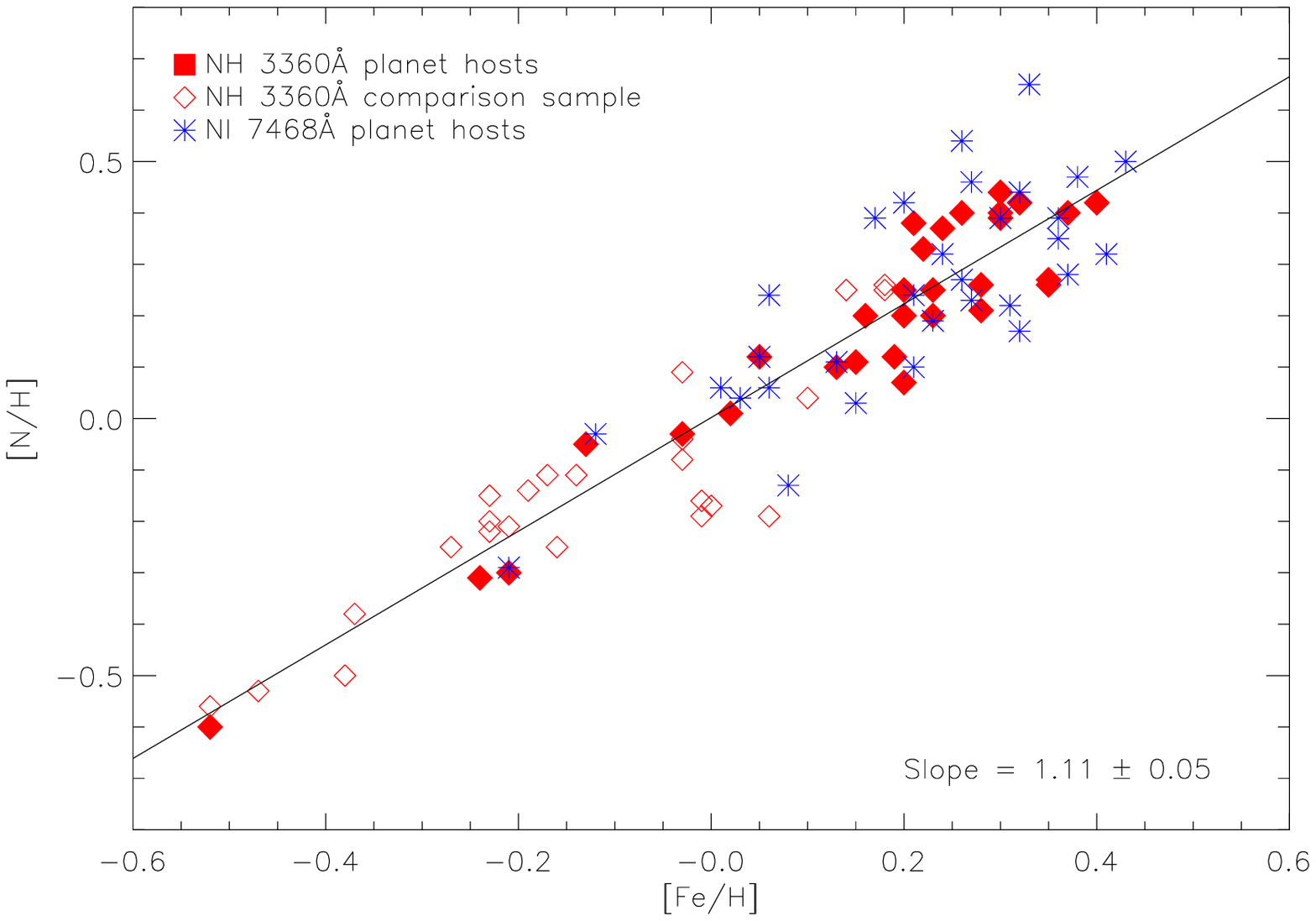}
\includegraphics[height=6cm]{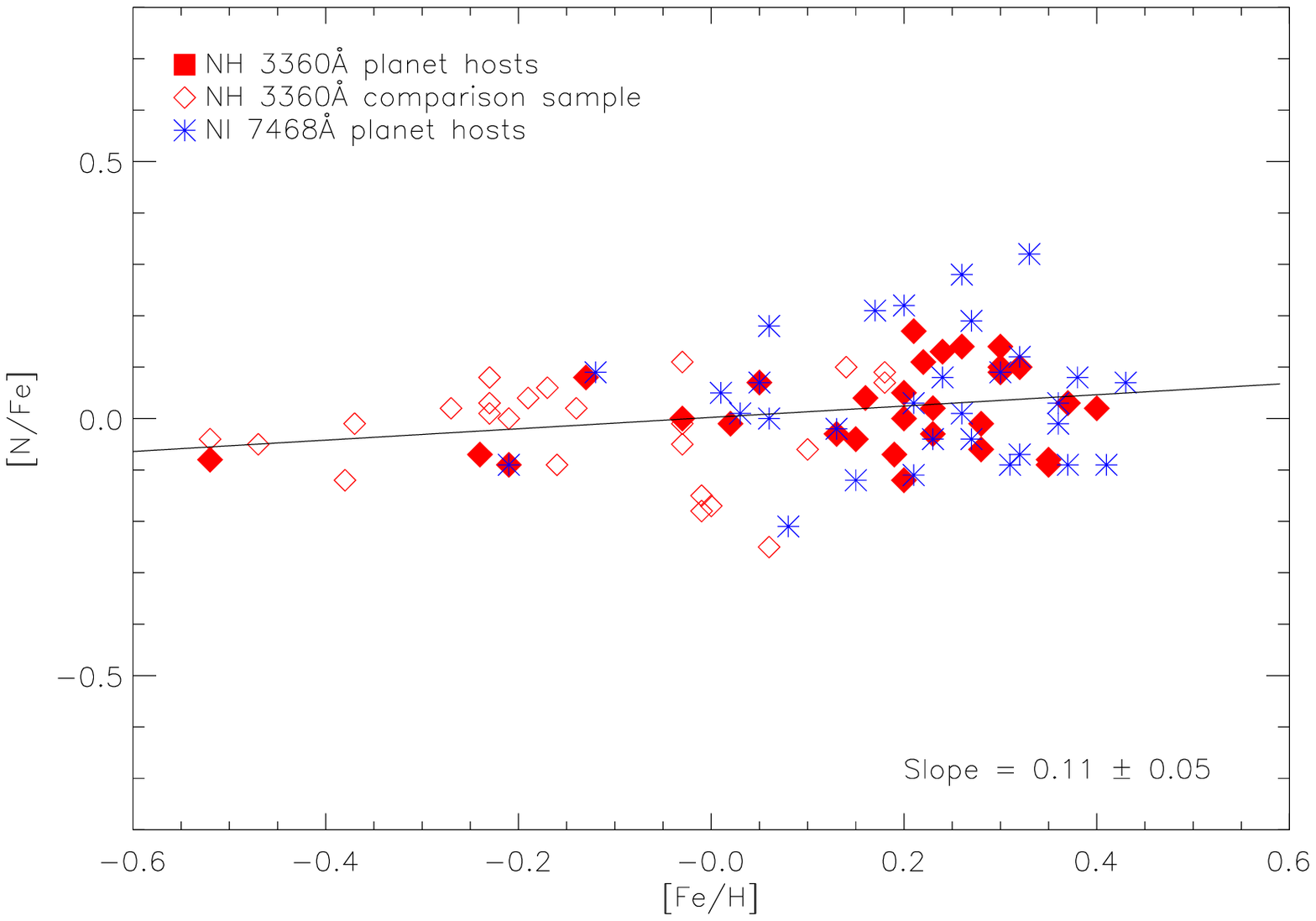}
\caption{[N/H] and [N/Fe] vs.\ [Fe/H] plots from NH 3360 \AA\ band and \ion{N}{i} 7468 \AA\ line values. Filled diamonds 
and open diamonds represent 
planet host and comparison sample stars from NH band synthesis, respectively, while
asterisks denote planet host stars from \ion{N}{i} line analysis. Linear least-squares fits to all samples together are 
represented and slope values are written at the bottom of each plot.}
\label{fig5}
\end{figure*}

\begin{table*}[t]
\caption[]{Comparison of nitrogen abundances determined from NH band synthesis and from \ion{N}{i} line analysis.}
\begin{scriptsize}
\begin{center}
\begin{tabular}{lcr}
\hline
\noalign{\smallskip}
Star &  [N/H]$_{\rm NH\,3360}$ & [N/H]$_{\rm \ion{N}{i}\,7468}$ \\ 
\noalign{\smallskip}
\hline
\hline 
\noalign{\smallskip}
\object{HD\,16141} & 0.11 & 0.03         \\ 
\object{HD\,19994} & 0.37 & 0.17         \\ 
\object{HD\,46375} & 0.07 & 0.42         \\ 
\object{HD\,120136} & 0.20 & 0.19        \\ 
\object{HD\,134987} & 0.40 & 0.39        \\ 
\object{HD\,143761} & $-$0.30 & $-$0.29  \\ 
\object{HD\,217107} & 0.40 & 0.28 \\ 
\object{HD\,222582} & 0.12 & 0.12 \\ 
\noalign{\smallskip}
\hline
\end{tabular}
\end{center}
\small{}
\end{scriptsize}
\label{tab7}
\end{table*}

\section{Nitrogen abundances from the \ion{N}{i} 7468 \AA\ line}

Several studies of nitrogen abundances in metal-rich dwarfs with planets using near-IR \ion{N}{i} lines have been published, 
but data for only a small number of planet host stars have been employed (fourteen planet-harbouring stars in the most 
extensive case, in
Takeda et al.\ \cite{Tak01}) and most of them do not even mention their nitrogen abundance results (Gonzalez \& Laws 
\cite{Gonz00}; Gonzalez et al.\ \cite{Gonz01}; Sadakane et al.\ \cite{Sad02}). Only Takeda et al.\ (\cite{Tak01}) 
have presented 
their nitrogen abundance results from the measurement of the \ion{N}{i} line at 8683 \AA, as well as abundance results 
for another 
four volatile elements  (C, O, S and Zn). They did not observe the tendency  predicted by the ``self-enrichment'' 
hypothesis ([$X$/H] $<$ [Fe/H] for metal-enriched planet host stars),  a weak inverse trend ([$X$/H] $>$ [Fe/H]) 
even seemed to 
appear for N and S.

We obtained nitrogen abundances for 31 planet host stars by equivalent width measurements of the \ion{N}{i} line at 
7468 \AA. All atmospheric parameters, EW values with uncertainties and abundance results are listed in Table~\ref{tab6}.
Figure~\ref{fig5} shows the [N/H] and [N/Fe] abundance ratios as functions of [Fe/H] for planet host stars, from NH band 
and \ion{N}{i} line sets, and for comparison sample stars. In Figure~\ref{fig4} (top lower panel), the 
[N/H] distributions are 
shown for both the planet host (with values from both indicators) and comparison star samples. In all the plots the added 
values 
follow the same behaviour as those resulting from NH band analysis. The previously commented trends (see
Section~\ref{NHtrend}) 
are confirmed.

\section{Comparison of the near-UV and 7468 \AA\ line results}

Two independent analyses were carried out, using different indicators of nitrogen abundance. We managed thereby to 
extend 
our data sample, as well as to check the agreement between both methods and to test the strength of our results. 

We had available to us two different sets of data. The first consisted of near-UV spectra of planet host 
and comparison sample stars. The other was composed of optical and near-IR spectra of stars with planets.  
The comparison between the results of both indicators was performed using the eight targets that both sets had in common.

Abundances resulting from both indicators are presented in Table~\ref{tab7}. We stress the remarkable agreement in most 
cases. In the case of \object{HD\,19994} and \object{HD\,217107}, both values fall in the uncertainty range resulting from 
the abundance analysis. Only for \object{HD\,46375} do the different indicators give different results.

\section{Reanalysis of published EW data}

\begin{table*}[t]
\caption[]{Nitrogen abundances calculated using EW values from other authors for a set of stars with planets
and brown dwarf companions.}
\begin{scriptsize}
\begin{center}
\begin{tabular}{lccccccr}
\hline
\noalign{\smallskip}
Star & Ref. & $T_\mathrm{eff}$ & $\log {g}$ & $\xi_t$ & [Fe/H] & EW & [N/H] \\       
     &      & (K)              & \, (cm\,s$^{-2}$) & \, (km\,s$^{-1}$) &     & (m\AA) &  \\
\noalign{\smallskip}
\hline 
\hline
\noalign{\smallskip}
\object{HD\,75289}  & 1  & $6143\pm53$ & $4.42\pm0.13$& $1.53\pm0.09$& $0.28\pm0.07$& 10.7 & 0.15     \\ 
\object{HD\,9826}   & 1  & $6212\pm64$ & $4.26\pm0.13$& $1.69\pm0.16$& $0.13\pm0.08$&  9.2 & $-$0.01  \\ 
\object{HD\,9826}   & 3  & $6212\pm64$ & $4.26\pm0.13$& $1.69\pm0.16$& $0.13\pm0.08$& 19.6 & 0.14     \\ 
\object{HD\,9826}   &avg.& $6212\pm64$ & $4.26\pm0.13$& $1.69\pm0.16$& $0.13\pm0.08$&  $-$   & 0.07   \\ 
\object{HD\,10697}  & 2  & $5641\pm28$ & $4.05\pm0.05$& $1.13\pm0.03$& $0.14\pm0.04$&  6.0 & 0.15     \\ 
\object{HD\,89744}  & 2  & $6234\pm45$ & $3.98\pm0.05$& $1.62\pm0.08$& $0.22\pm0.05$& 13.3 & 0.09     \\ 
\object{HD\,89744}  & 3  & $6234\pm45$ & $3.98\pm0.05$& $1.62\pm0.08$& $0.22\pm0.05$& 22.9 & 0.12     \\ 
\object{HD\,89744} &avg.& $6234\pm45$ & $3.98\pm0.05$& $1.62\pm0.08$& $0.22\pm0.05$&  $-$   & 0.11    \\ 
\object{HD\,217014} & 2  & $5804\pm36$ & $4.42\pm0.07$& $1.20\pm0.05$& $0.20\pm0.05$&  9.0 & 0.34     \\ 
\object{HD\,117176} & 3  & $5560\pm34$ & $4.07\pm0.05$& $1.18\pm0.05$& $-0.06\pm0.05$&  4.0 & $-$0.21 \\ 
\object{HD\,106252} & 4  & $5899\pm35$ & $4.34\pm0.07$& $1.08\pm0.06$& $-0.01\pm0.05$&  4.9 & $-$0.33 \\ 
\object{HD\,134987} & 2  & $5776\pm29$ & $4.36\pm0.07$& $1.09\pm0.04$& $0.30\pm0.04$&  10.6 & 0.43    \\ 
\object{HD\,143761} &  3 & $5853\pm25$ & $4.41\pm0.15$& $1.35\pm0.07$& $-0.21\pm0.04$&  6.1 & $-$0.12 \\ 
\object{HD\,168443} &  2 & $5617\pm35$ & $4.22\pm0.05$& $1.21\pm0.05$& $0.06\pm0.05$&  4.8 & 0.12     \\ 
\object{HD\,186427} &  3 & $5772\pm25$ & $4.40\pm0.07$& $1.07\pm0.04$& $0.08\pm0.04$&  10.8 & 0.22    \\ 
\object{HD\,187123} &  4 & $5845\pm22$ & $4.42\pm0.07$& $1.10\pm0.03$& $0.13\pm0.03$&   6.5 & $-$0.14 \\ 
\object{HD\,95128} &  3 & $5954\pm25$ & $4.44\pm0.10$& $1.30\pm0.04$& $0.06\pm0.03$&  13.0 & 0.18     \\ 
\object{HD\,12661} &  2 & $5702\pm36$ & $4.33\pm0.08$& $1.05\pm0.04$& $0.36\pm0.05$&  9.0 & 0.38      \\ 
\object{HD\,217107} & 2  & $5663\pm36$ & $4.34\pm0.08$& $1.11\pm0.04$& $0.37\pm0.05$&  3.9 & $-$0.03  \\ 
\object{HD\,222582} & 2  & $5843\pm38$ & $4.45\pm0.07$& $1.03\pm0.06$& $0.05\pm0.05$&  4.9 & 0.02     \\ 
\object{HD\,75732} &  3 & $5279\pm62$ & $4.37\pm0.18$& $0.98\pm0.07$& $0.33\pm0.07$& 10.5 & 0.70      \\ 
\object{HD\,120136} & 1  & $6339\pm73$ & $4.19\pm0.10$& $1.70\pm0.16$& $0.23\pm0.07$& 16.7 & 0.21     \\ 
\object{HD\,130322} & 4  & $5392\pm36$ & $4.48\pm0.06$& $0.85\pm0.05$& $0.03\pm0.04$&  3.2 & $-$0.03  \\ 
\object{HD\,141937} & 4  & $5909\pm39$ & $4.51\pm0.08$& $1.13\pm0.06$& $0.10\pm0.05$&  7.7 & $-$0.06  \\ 
\object{HD\,168746} & 4  & $5601\pm33$ & $4.41\pm0.12$& $0.99\pm0.05$& $-0.08\pm0.05$&  5.6 & 0.04    \\ 
\object{HD\,190228} & 4  & $5327\pm35$ & $3.90\pm0.07$& $1.11\pm0.05$& $-0.26\pm0.06$&  3.5 & $-$0.07 \\ 
\noalign{\smallskip}
\hline
\end{tabular}
\end{center}
\footnotetext{}{Ref.1 from Gonzalez \& Laws (\cite{Gonz00})}\\
\footnotetext{}{Ref.2 from Gonzalez et al.\ (\cite{Gonz01})}\\
\footnotetext{}{Ref.3 from Takeda et al.\ (\cite{Tak01})}\\
\footnotetext{}{Ref.4 from Sadakane et al.\ (\cite{Sad02})}\\
\end{scriptsize}
\label{tab8}
\end{table*}

\begin{figure*}
\centering
\includegraphics[height=6cm]{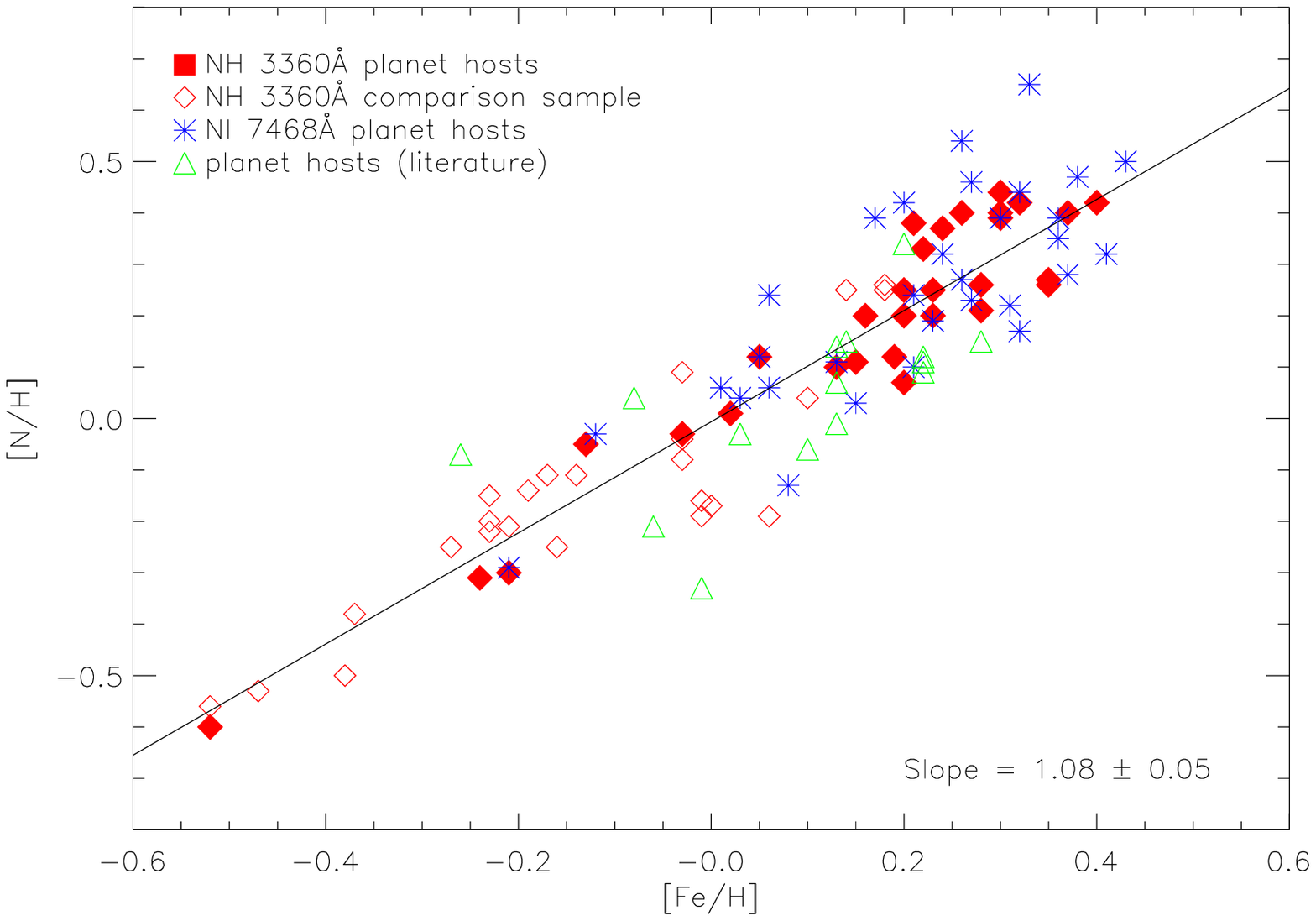}
\includegraphics[height=6cm]{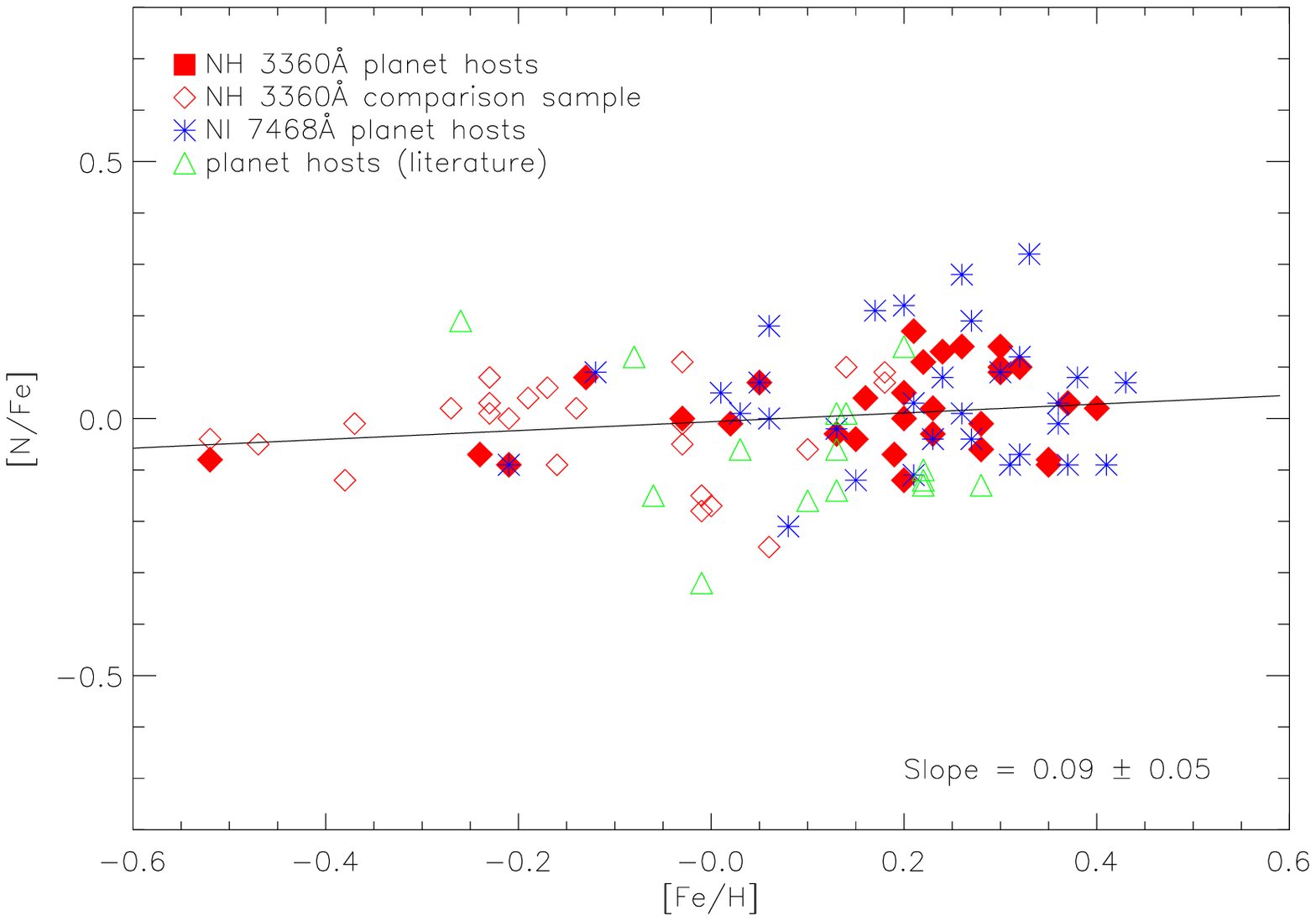}
\caption{[N/H] and [N/Fe] vs.\ [Fe/H] plots from NH 3360 \AA\ band, \ion{N}{i} 7468 \AA\ line and literature values. Filled 
and open diamonds represent planet host and comparison sample stars from NH band synthesis, respectively.
Asterisks and open triangles denote planet host stars from \ion{N}{i} line and from literature
values, respectively. Linear least-squares fits to all samples together are represented and slope values are written at the 
bottom of each 
plot.}
\label{fig6}
\end{figure*}

\begin{table*}[t]
\caption[]{Comparison of nitrogen abundances determined using EW values (in m\AA) from our data and from other authors. }
\begin{scriptsize}
\begin{center}
\begin{tabular}{lcccccccccr}
\hline
\noalign{\smallskip}
Star & $T_\mathrm{eff}$ & $\log {g}$       & $\xi_t$   & [Fe/H]  & EW$^1$ & EW$^2$ & EW$^3$ & [N/H]$^1$ & [N/H]$^2$ & [N/H]$^3$ \\
     & (K)              &  (cm\,s$^{-2}$) & (km\,s$^{-1}$) &    & 7468 & 7468 & 8683 &   &  &  \\
\noalign{\smallskip}
\hline
\hline
\noalign{\smallskip}
\object{HD\,134987} & 5776 & 4.36 & 1.09 & 0.30 & 10.0 & 10.6 & $-$ & 0.39 & 0.43 & $-$ \\ 
\object{HD\,143761} & 5853 & 4.41 & 1.35 & $-$0.21 & 2.5 & $-$ & 6.1 & $-$0.29 & $-$ & $-0.12$ \\
\object{HD\,168443} & 5617 & 4.22 & 1.21 & 0.06 & 6.0 & 4.8 & $-$ & 0.24 & 0.12 & $-$ \\
\object{HD\,186427} & 5772 & 4.40 & 1.07 & 0.08 & 3.4 & $-$ & 10.8 & $-$0.13 & $-$ & 0.22 \\
\object{HD\,187123} & 5845 & 4.42 & 1.10 & 0.13 & 6.1 & $-$ & 6.5 & 0.11 & $-$ & $-0.14$ \\
\object{HD\,95128} & 5954 & 4.44 & 1.30 & 0.06 & 6.4 & $-$  & 13.0 & 0.06 & $-$ & 0.18 \\
\object{HD\,12661} & 5702 & 4.33 & 1.05 & 0.36 & 8.5 & 9.0 & $-$ & 0.35 & 0.38 & $-$ \\
\object{HD\,217107} & 5663 & 4.34 & 1.11 & 0.37 & 8.7 & 3.9 & $-$ & 0.28 & $-0.03$ & $-$\\
\object{HD\,222582} & 5843 & 4.45 & 1.03 & 0.05 & 6.0 & 4.9 & $-$ & 0.12 & 0.02 & $-$ \\
\object{HD\,75732} & 5279 & 4.37 & 0.98 & 0.33 & 6.5 & $-$ & 10.5 & 0.65 & $-$ & 0.70 \\
\object{HD\,120136} & 6339 & 4.19 & 1.70 & 0.23 & 16.2 & 16.7 & $-$ & 0.19 & 0.21 & $-$ \\
\noalign{\smallskip}
\hline
\end{tabular}
\end{center}
{$^1$ EW values from our data.\\ $^2$ EW values from
Gonzalez \& Laws (\cite{Gonz00}) and Gonzalez et al.\ (\cite{Gonz01}).\\ $^3$ EW values from Takeda et al.\
(\cite{Tak01}) and Sadakane et al.\ (\cite{Sad02}). }
\end{scriptsize}
\label{tab9}
\end{table*}
Several recent studies have presented measurements of \ion{N}{i} lines in targets with known planets 
(Gonzalez \& Laws \cite{Gonz00}; 
Gonzalez et al.\ \cite{Gonz01}; Takeda et al.\ \cite{Tak01}; Sadakane et al.\ \cite{Sad02}). We have collected all the
EW values 
determined in these papers and used them to calculate nitrogen abundances with atmospheric parameters from Santos et al.\  
(\cite{San03b}b).   

Gonzalez \& Laws (\cite{Gonz00}) and Gonzalez et al.\ (\cite{Gonz01}) measured \ion{N}{i} line at 7468 \AA, as we did;
therefore, we could compare EW values in some of the targets that both analyses had in common. 
Takeda et al.\ (\cite{Tak01}) and Sadakane 
et al.\ (\cite{Sad02}) used the \ion{N}{i} line at 8683 \AA, so we were not able to compare our EW values with theirs,
 but only [N/H].
Besides allowing us to carry out a comparison, the literature values added new abundance measures to our data sample.

This reanalysis was carried out in the same way described in Section~\ref{NIan}. In the case of the \ion{N}{i} line at 
8683 \AA, the solar gf value computed using the equivalent width (7.0 m\AA) measured in the Solar Atlas (Kurucz et al.\ 
\cite{Kur84}) was $\log {gf}=0.068$.  
All the EW values from the literature, the atmospheric parameters from Santos et al.\ (\cite{San03b}b) and the resulting
nitrogen abundances are presented in Table~\ref{tab8}. All the results for the targets in common with our data 
sample are listed 
in Table~\ref{tab9}. We note that the results from different sources are in good agreement. Significant differences exist 
only 
for 
\object{HD\,186427}, \object{HD\,187123} and \object{HD\,217107}.
The [N/H] and [N/Fe] vs.\ [Fe/H] trends from the literature EW values and from our data are represented in Figure~\ref{fig6}. 
Figure~\ref{fig4} (bottom panel) shows the [N/H] distribution for comparison sample and planet host stars from all
the sources.
We note that the new values confirm the trends discussed in the previous sections (see Section~\ref{NHtrend}).

\section{Discussion and Conclusions}
\label{Disc}

In the present study we have determined the nitrogen abundances in 51 planet host stars and in 25 stars from a 
comparison sample
with no known planetary mass companion. Among the set of planet hosts, 28 targets have been analysed by spectral synthesis
of the NH band at 3360 \AA, while equivalent width measurements of the \ion{N}{i} 7468.27 \AA\
were carried out for 31 stars. Another 15 planet hosts 
were subsequently added to our study, with EW values taken from other authors' works. Moreover, we were able to 
check that the different indicators are in remarkable agreement for several common targets. This result is an independent 
and
homogeneous analysis of nitrogen abundances in 91 solar-type dwarfs, 66 with planets and 25 from a volume-limited comparison 
sample.

The behaviour of volatiles in planet-harbouring stars with regard to a comparison sample can be very informative
for checking the ``self-enrichment'' hypothesis. If the accretion of metal-enriched planetary material were a key parameter 
for the observed enhancement of iron abundances, then volatile abundances in planet host stars would not show as much
overabundance as refractories do since elements with a lower condensation temperature are expected to be deficient in
accreted materials. In that case, the following trend would be observed in planet host stars: [N/H] $<$ [Fe/H].

Our results suggest that planet hosts stars do not behave in such a way. The [N/H] vs.\ [Fe/H] plots show that nitrogen 
abundance scales with that of the iron. In the [N/Fe] vs.\ [Fe/H] plots, no significant trace of [N/Fe] $<$ 0 appears in 
stars 
with planets. Moreover, planet host and comparison sample stars have the same behaviour. Trends from either sample are 
quite indistinguishable. This emphasizes that the trend predicted from the ``self-enrichment'' scenario is not observed at 
all. 

Previous studies have obtained similar results for other volatiles in some planet-harbouring stars. Gonzalez et al.\
 (\cite{Gonz01}) 
corrected the low [C/Fe] values found in a previous study (Gonzalez \& Laws \cite{Gonz00}) and concluded that [C/Fe] and 
[O/Fe] in planet hosts do not display significant differences from those in field dwarfs of the same [Fe/H]. Takeda et al.\ 
(\cite{Tak01}) found the same trend as we did for N, C, S, O and Zn in 14 planet hosts, and in Sadakane et al.\ 
(\cite{Sad02}) this result was confirmed for C and O in another 12 planet-harbouring stars. Our study confirms these 
results for a significantly larger set of stars with planets, as well as in a comparison sample of stars with no discovered 
planetary-mass companion. This seems to support a scenario in which the formation of planets is particularly dependent on the 
high metallicity of the primordial cloud.

In the future, it will be important to pursue further uniform studies of volatile element (C, O, S and Zn) abundances 
in planet 
hosts and stars with no known planets. Well-defined [$X$/Fe] vs.\ [Fe/H] distributions can provide us with
precise information
 in 
order to confirm or discard hypotheses about planetary formation.

Another important piece of information that emerges from nitrogen abundance trends concerns  nitrogen sources. The
latest studies propose two sources of primary nitrogen: intermediate mass (4--8
$M_{\sun}$) and massive stars ($M>8\ M_{\sun}$). If the latter source were the
main contributor to nitrogen at high [Fe/H], N and Fe abundances 
would appear to be uncoupled and nitrogen would be overabundant relative to iron.
Our study of nitrogen abundances in 91 solar-type dwarfs suggests that nitrogen keeps pace with iron. The [N/H]
vs.\ [Fe/H] plots show that both elements behave quite similarly, and the [N/Fe] vs.\ [Fe/H] plots 
indicate that no significant
nitrogen overabundance exists relative to iron in the metallicity range $-0.6<$ [Fe/H] $<+0.4$.
We also note that Shi et al.\ (\cite{Shi02}) recently obtained that [N/Fe] is about solar in a sample of 90 disk stars; therefore, they
concluded that nitrogen is produced mostly by intermediate mass stars.

Nevertheless, secondary production due to ILMS is the predominant effect on the [N/Fe] curve around solar
metallicities. Only the [N/Fe] trend at [Fe/H] $<-1.0$ can provide conclusive arguments about primary sources. 
At this time, we are extending our study of [N/H] ratios to lower metallicities, by applying the same analysis of the 
NH band 
at 
3360 \AA\ to a sample of metal-poor dwarfs. The results will be very informative in the framework of investigating
the sources of primary nitrogen in the early Galaxy.
  
\begin{acknowledgements}
We would like to thank J. I. Gonz\'alez Hern\'andez for kindly providing us with his program FITTING. 
We wish to thank Dr. G. Meynet (Observatoire de Gen\`eve, Switzerland) for many fruitful discussions and his comments on the 
text.
{ The anonymous referee is thanked for many useful suggestions and comments.} 
IRAF is distributed by the National Optical Astronomy Observatories, operated by the Association of Universities for 
Research in Astronomy, Inc., under contract with the National Science Foundation, USA.
\end{acknowledgements}

%---------------------------bibliography---------------------------

\listofobjects 
  
\end{document}